\documentstyle[equation,12pt]{article}
\def\lsim{<\kern-2.5ex\lower0.85ex\hbox{$\sim$}\ }
\def\rsim{>\kern-2.5ex\lower0.85ex\hbox{$\sim$}\ }
\def\ni{\noindent}
\def\LAMBDABAR
{\hbox{$\lambda$\kern-0.52em\raise+0.45ex\hbox{--}\kern+0.2em}}

\begin{document}

\begin{center}
{\bf GLOBAL SYMMETRIES OF TIME-DEPENDENT SCHR\"ODINGER EQUATIONS}
\end{center}

\begin{center}
Susumu Okubo\\ Department of Physics and Astronomy \\ University
of Rochester \\ Rochester, NY 14627, USA
\end{center}

\begin{abstract}
Some  symmetries of time-dependent Schr\"odinger equations for
inverse quadratic, linear, and quadratic potentials have been
systematically examined by using a method suitable to the problem.
Especially, the symmetry group for \hfill the \hfill case \hfill
of \hfill the \hfill linear \hfill potential \hfill turns \hfill
out \hfill to \hfill be \hfill a \hfill semi-direct \hfill product

\vspace{-.41in}

\noindent $SL(2,R) \ \ \ \,
\begin{picture}(40,40) \put(-1,3){\circle{11}}\hspace{-7pt} s
\end{picture} \!\!\!\!\!\!\!\!\!\!\!\!\!\! T_2(R)$ of the $SL(2,R)$ with a
 two-dimensional real translation group $T_2 (R)$. Here, the time
 variable $t$ transforms as $t \rightarrow t^\prime =
 (ct+d)/(at+b)$ for real constants $a,\ b,\ c$, and $d$ satisfying
 $bc - ad =1$ with an accompanying transformation for the space
 coordinate $x$.
\end{abstract}

\section{Formulation}
\setcounter{equation}{0}

 Many solutions of Schr\"odinger equations
are known$^{1-4)}$ (and earlier references quoted therein) to
possess dynamical (or hidden) symmetries which are not apparent at
first glance.

In a different approach, Eastwood$^{5)}$ in his study of symmetry
of Laplace equation has observed the following: Suppose that a
pair of functions $U({\bf x} , \mbox{\boldmath $\partial$})$ and
$W ({\bf x}, \mbox{\boldmath $\partial$})$ of the coordinate ${\bf
x}$ and its derivative $\mbox{\boldmath $\partial$}$ satisfies

$$W ({\bf x} , \mbox{\boldmath $\partial$} ) \Delta = \Delta U({\bf x},
\mbox{\boldmath $\partial$}) \ ,\ \left( \Delta = \sum^n_{j=1} \
{\partial^2 \over
\partial x_j^2}\right)$$

\ni for the Laplacian $\Delta$. Then, if $\psi ({\bf x})$ is a
solution of the Laplace equation $\Delta \psi ({\bf x} ) =0$, then
so will be

$$\psi^\prime ({\bf x}) = U ({\bf x},\mbox{\boldmath $\partial$}) \psi
({\bf x}) \quad .$$

The purpose of this note is to utilize an analogous method to
systematically find global symmetries of time-independent
Schr\"odinger equations. We consider the equation of motion of the
form:

\begin{equation}
{\partial \over \partial t} \ \psi (t, {\bf x} ) = k \{ \Delta - V
({\bf x}) \} \psi (t, {\bf x}) \quad . \label{eq:oneone}
\end{equation}

\ni If $k$ is purely imaginary with $k= -i\hbar /2m$, then it
describes the standard Schr\"odinger equation, while the case of
$k$ being real implies a diffusion-type equation. In what follows,
we will  always assume that the parameter $k$ is either real or
purely imaginary with $t$ and ${\bf x}$ being real unless it is
stated otherwise.

Consider now a vector space of all suitably smooth functions of
$t$ and ${\bf x}$, which may be complex. Suppose that we can find
a pair of linear operators $U$ and $W$ in this space to satisfy
the condition

\begin{equation}
W \left\{ {\partial \over \partial t} -k \Delta + k V ({\bf x}
)\right\} = \left\{ {\partial \over \partial t} - k \Delta + k V
({\bf x})\right\} U \quad . \label{eq:onetwo}
\end{equation}

\ni Then, if $\psi (t, {\bf x})$ satisfies Eq.(\ref{eq:oneone}),
the new function given by

\begin{equation}
\psi^\prime (t, {\bf x}) = (U \psi) (t, {\bf x}) \equiv U \psi (t,
{\bf x}) \label{eq:onethree}
\end{equation}

\ni will also obey the same relation, i.e. we have

\begin{equation}
{\partial \over \partial t}\ \psi^\prime (t, {\bf x}) = k \{
\Delta - V ({\bf x})\} \psi^\prime (t, {\bf x}) \quad .
\label{eq:onefour}
\end{equation}

Some explicit forms of $U$ and $W$ can be found as follows. Let us
consider the coordinate transformation of form for ${\bf x} =
\left( x_1 , x_2 , \dots , x_n \right)$,

\begin{subequations} \label{foo}
\begin{eqnarray}
t \rightarrow t^\prime &=& \phi (t)\quad ,  \label{foo-a}
 \label{eq:onefive}\\
\noalign{\vskip 4pt}%
x_j \rightarrow x^\prime_j &=& F_j (t, {\bf x} ) \quad , \quad (j
= 1,2,\dots , n) \label{foo-b} \label{eq:onefive}
\end{eqnarray}
\end{subequations}

\ni for some differentiable functions $\phi (t)$ and $F_j (t, {\bf
x})$ to be determined. The action of the linear operator $U$  to a
function $\psi (t, {\bf x})$ is then assumed to be given as a
multiplication of a function after the coordinate transformation,
i.e.,

\begin{equation}
\psi^\prime (t, {\bf x}) = U \psi (t, {\bf x}) = K (t, {\bf x})
\psi (t^\prime , {\bf x}^\prime ) \label{eq:onesix}
\end{equation}

\ni where $K(t, {\bf x})$ is a function of $t$ and ${\bf x}$ to be
determined. When we note

\begin{eqnarray*}
{\partial \over \partial t} &=& \dot \phi (t) \ {\partial \over
\partial t^\prime} + \sum^n_{j=1} \ {\partial F_j (t,
{\bf x}) \over \partial t}\ {\partial \over \partial x_j^\prime}\\
\noalign{\vskip 4pt}%
{\partial \over \partial x_j} &=& \sum^n_{k=1}\ {\partial F_k (t,
{\bf x}) \over \partial x_j}\ {\partial \over \partial x^\prime_k}
\end{eqnarray*}

\ni then we calculate

\begin{eqnarray}
& & \left\{ {\partial \over \partial t} - k \Delta + k V({\bf
x})\right\} \psi^\prime (t, {\bf x}) = \left\{ {\partial \over
\partial t} - k \Delta + kV ({\bf x}) \right\} \left[ K(t, {\bf x}) \psi (t^\prime ,
 {\bf x}^\prime )\right] \nonumber\\
\noalign{\vskip 4pt}%
& & \ = \left\{ \left[ {\partial \over \partial t} - k \Delta + kV
({\bf x}) \right] K(t, {\bf x})\right\} \psi (t^\prime , {\bf
x}^\prime ) + \nonumber\\
\noalign{\vskip 4pt}%
& &\ \ \ + \ K (t, {\bf x}) \sum^n_{j=1} \left\{ {\partial F_j (t,
{\bf x}) \over \partial t} - k \Delta F_j (t, {\bf x}) \right.
\nonumber\\
\noalign{\vskip 4pt}%
& & \ \ \ \left. -\  2 k \sum^n_{\ell =1} \ {\partial \log K(t,
{\bf x}) \over \partial x_\ell} \ {\partial F_j (t, {\bf x}) \over
\partial x_\ell} \right\}\ {\partial \over
\partial x^\prime_j}\ \psi (t^\prime , {\bf x}^\prime
)\nonumber\\
\noalign{\vskip 4pt}%
& &\ \ \ +\  K(t, {\bf x}) \left\{ \dot \phi (t) \ {\partial \over
\partial t^\prime} - k \sum^n_{j, \ell=1} \left( \sum^n_{i=1}\
{\partial F_j \over \partial x_i}\ {\partial F_\ell \over \partial
x_i} \right) \ {\partial^2 \over \partial x^\prime_j \partial
x^\prime_\ell} \right\} \psi (t^\prime , {\bf x}^\prime ) \quad .
\label{eq:oneseven}
\end{eqnarray}

\ni If $F_j (t, {\bf x})$, and $K (t, {\bf x})$ satisfy relations:

\begin{subequations} \label{foo}
\begin{eqnarray}
&{\rm (i)}&\quad \sum^n_{i=1} {\partial F_j (t, {\bf x}) \over
\partial x_i}  \  {\partial F_\ell (t, {\bf x}) \over \partial
x_i}= \delta_{j \ell} \dot \phi (t) \left( \equiv \delta_{j \ell}\
{d \over
dt}\ \phi (t) \right)\quad , \label{foo-a} \label{eq:oneeighta}\\
\noalign{\vskip 8pt}%
&{\rm (ii)}&\quad \left( {\partial \over \partial t} - k \Delta
\right) F_j (t, {\bf x})
 = 2k \sum^n_{\ell =1} \ {\partial \log
 K(t, {\bf x})
 \over \partial x_\ell }\ {\partial F_j
 (t, {\bf x}) \over \partial x_\ell} \quad ,
\label{foo-b} \label{eq:oneeightb}\\
 \noalign{\vskip 8pt}%
&{\rm (iii)}&\quad \left\{ {\partial \over \partial t} - k \Delta
+ k V ({\bf x}) \right\} K(t, {\bf x}) = k \dot \phi (t) V({\bf
x}^\prime) K(t, {\bf x}) \quad , \label{foo-c}
\label{eq:oneeightc}
\end{eqnarray}
\end{subequations}

\ni then Eq. (\ref{eq:oneseven}) becomes

\begin{equation}
\left\{ {\partial \over \partial t} - k \Delta + kV ({\bf
x})\right\} \psi^\prime (t, {\bf x}) = \dot \phi (t) K(t, {\bf x})
\left\{ {\partial \over
\partial t^\prime} - k \Delta^\prime + k V ({\bf x}^\prime)
\right\} \psi ( t^\prime, {\bf x}^\prime ) \quad ,
\label{eq:onenine}
\end{equation}

\ni which reproduces Eq. (\ref{eq:onetwo}) with actions of $U$ and
$W$ given by

\begin{subequations} \label{foo}
\begin{eqnarray}
U \psi (t, {\bf x}) &=& K(t, {\bf x}) \psi (t^\prime , {\bf
x}^\prime) \label{foo-a}\label{eq:oneten}\\
\noalign{\vskip 4pt}%
W \hat \psi (t, {\bf x}) &=& \dot \phi (t) K(t, {\bf x}) \hat \psi
(t^\prime , {\bf x}^\prime )  \label{foo-b}\label{oneten}
\end{eqnarray}
\end{subequations}

\ni for any two functions $\psi$ and $\hat \psi$. For the present
problem, we have

$$ \hat \psi (t, {\bf x}) = \left( {\partial \over \partial t} -k
\Delta+ k V ({\bf x}) \right) \psi (t, {\bf x}) \quad .$$

\ni Summarizing, we have proved the following Theorem.

\bigskip

\ni \underbar{\bf Theorem 1.1}

\medskip

 Let functions $\phi (t)$, $F_j (t, {\bf x})$, and
$K(t, {\bf x})$ satisfy Eqs. (1.8) with $t^\prime$ and
$x^\prime_j$ being given by Eqs. (1.5). Then for any function
$\psi (t, {\bf x})$ which satisfies Eq. (\ref{eq:oneone}), i.e.,

$${\partial \over \partial t}\ \psi (t, {\bf x}) = k \left\{
\Delta -V({\bf x}) \right\} \psi (t, {\bf x}) \quad ,$$

\ni the new function given by

$$\psi^\prime (t, {\bf x}) = K (t, {\bf x}) \psi (t^\prime , {\bf
x}^\prime) = K(t, {\bf x}) \psi \left( \phi (t),F_j (t, {\bf x})
\right)$$

\ni is also a solution of the same generalized Schr\"odinger
equation Eq. (\ref{eq:oneone}).

\bigskip

\ni \underbar{\bf Remark 1.2}

\medskip

We may call the pair of linear operators, $(U,W)$ satisfying Eq.
(\ref{eq:onetwo}) be admissible. Then, for the second admissible
pair $(U^\prime , W^\prime)$ their product $(UU^\prime ,
WW^\prime)$ is clearly also admissible. Moreover, the special pair
$(1,1)$ acts as the identity. Therefore, a set of all admissible
pairs form a semi-group. If the pair is invertible, then they
present a symmetry group of Eq. (\ref{eq:oneone}). More
explicitly, if the transformation with certain $V({\bf x})$ are
chosen as in Eq. (1.16), one gets a set $S$ of possible $\phi$,
$F$, and $K$ with elements $s\ \epsilon \ S$ parametrized by real
numbers $a$, $b$, $\dots$, i.e., $s = s(a,b,\dots)$ and a set of
solutions $\psi (t, {\bf x}; a, b, \dots)$.  It will be shown in
subsequent sections that the $s\  \epsilon\  S$ and hence the
corresponding solutions can be transformed into each other
formally via e.g., $SL(2,R)$ depending upon $V({\bf x})$.

\bigskip

We can modify Theorem 1.1 slightly as follows. Let $V_0 ({\bf x})$
and $V( {\bf x})$ be two potentials, and suppose that the pair of
linear operators $(U,W)$ now satisfies

\begin{equation}
W \left\{ {\partial \over \partial t} - k \Delta + kV_0 ({\bf
x})\right\}  = \left\{ {\partial \over \partial t} -k \Delta + k
V({\bf x}) \right\} U \label{eq:oneeleven}
\end{equation}

\ni instead Eq. (\ref{eq:onetwo}). We can proceed exactly in the
same way as in the previous case, and we prove the following
theorem.

\bigskip

\ni \underbar{\bf Theorem 1.3}

\medskip

Let functions $\phi (t),\ F_j (t, {\bf x})$ and $K (t,{\bf x})$ as
in Theorem 1.1 except that Eq. (\ref{eq:oneeightc}) is now
replaced by

\begin{equation}
\left\{ {\partial \over \partial t} - k\Delta + k V({\bf x})
\right\} K(t, {\bf x}) = k \dot \phi (t) V_0 ({\bf x}^\prime) K(t,
{\bf x}) \quad . \label{eq:onetwelve}
\end{equation}

\ni Then, for any function $\psi_0 (t,{\bf x})$ satisfying

\begin{equation}
\left\{ {\partial \over \partial t} -k\Delta + k V_0 ({\bf x})
\right\} \psi_0 (t,{\bf x} ) = 0 \quad , \label{eq:onethirteen}
\end{equation}

\ni the function given by

\begin{equation}
\psi (t, {\bf x}) = U \psi_0 (t,{\bf x}) = K (t, {\bf x}) \psi_0
(t^\prime , {\bf x}^\prime) \label{eq:onefourteen}
\end{equation}

\ni satisfies

\begin{equation}
\left\{ {\partial \over \partial t} - k \Delta + kV ({\bf x})
\right\} \psi (t, {\bf x}) = 0 \quad . \quad \vrule height 0.9ex
width 0.8ex depth -0.1ex
\end{equation}

\bigskip

\ni \underbar{\bf Remark 1.4}

\medskip

Examples satisfying Theorem 1.3 for $V_0 ({\bf x}) =0$ will be
given in sections 3 and 4. If $U^{-1}$ and $W^{-1}$ exist, then we
can conversely express $\psi_0 (t, {\bf x})$ in terms of $\psi (t,
{\bf x})$.\ $\vrule height 0.9ex width 0.8ex depth -0.1ex$

\bigskip

Returning now to the original discussion of Theorem 1.1, it is in
general difficult to find solutions of differential equations,
Eqs. (1.8). However, for three cases of $V({\bf x})$ being inverse
square, linear, and quadratic potentials, we can solve them as
follows. The explicit forms of $F_j (t, {\bf x})$ and $K(t, {\bf
x})$ can then be assumed to be

\begin{subequations} \label{foo}
\begin{eqnarray}
t^\prime &=& \phi (t) \quad , \label{foo-a} \label{eq:onesixteena}\\
\noalign{\vskip 4pt}%
x^\prime_j &=& F_j (t, {\bf x}) = \xi (t) x_j + f_j (t) \quad ,
\label{foo-b} \label{eq:onesixteenb}\\
 \noalign{\vskip 4pt}%
K(t, {\bf x}) &=& \exp \left\{ A (t) + \sum^n_{j=1} B_j (t) x_j +
C (t) {\bf x}^2 \right\} \quad , \label{foo-c}
\label{eq:onesixteenc}
\end{eqnarray}
\end{subequations}

\ni for some functions $\phi (t)$, $\xi (t)$, $f_j (t)$, $A(t)$,
$B_j (t)$, and $C(t)$ of $t$ to be determined as in the following
Proposition.

\bigskip

\ni \underbar{\bf Proposition 1.5}

\medskip

Eqs. (1.8) will be satisfied for the ans\"atz Eqs. (1.16), if we
have

\begin{subequations}\label{foo}
\begin{eqnarray}
&{\rm (i)}&\quad \dot \phi (t) = \xi^2 (t) \label{foo-a}
\label{eq:oneseventeena}\\
\noalign{\vskip 8pt}%
&{\rm (ii)}&\quad B_j (t) = {1 \over 2k}\ {\dot f_j (t) \over \xi
(t)} \label{foo-b} \label{eq:oneseventeenb}\\
 \noalign{\vskip 8pt}%
&{\rm (iii)}& \quad C(t) = {1 \over 4k}\ {\dot \xi (t) \over \xi
(t)} \quad , \label{foo-c}\label{eq:oneseventeenc}
\end{eqnarray}
\end{subequations}

\ni provided that $K(t, {\bf x})$ satisfies Eq. (1.8c), i.e.,

\begin{equation}
\left\{ {\partial \over \partial t} -k\Delta +kV ({\bf x})
\right\} K(t,{\bf x}) = k \dot \phi (t) V ({\bf x}^\prime) K(t,
{\bf x}) \quad . \quad \vrule height 0.9ex width 0.8ex depth
-0.1ex\label{eq:oneeighteen}
\end{equation}

We will solve these differential equations, Eqs. (1.17) and
(\ref{eq:oneeighteen}) in subsequent sections. The simplest case
of the inverse quadratic potential will be discussed in section 2,
where the symmetry group is $SL(2,R)$. On the contrast, the
symmetry of

\vspace{-.41in}

\ni the linear potential  is a larger one of  $SL(2,R) \ \ \ \,
\begin{picture}(40,40)
\put(-1,3){\circle{11}}\hspace{-7pt} s
\end{picture}
 \!\!\!\!\!\!\!\!\!\!\!\!\!\! T_2(R)$  which is the semi-direct product
 of $SL(2,R)$ with a real two-dimensional translation group
 $T_2(R)$. This will be presented in section 3. For the case of
 the quadratic potential in section 4, the

\vspace{-.41in}

\ni  symmetry is now either a semi-group or a sub-group of
$SL(2,C) \ \ \ \,
\begin{picture}(40,40)
\put(-1,3){\circle{11}}\hspace{-7pt} s
\end{picture}
 \!\!\!\!\!\!\!\!\!\!\!\!\!\! T_2(C)$, depending upon choices of
 parameters involved in theory. In section 5, we will discuss the
 associated Lie algebras of these groups.

Last, we simply remark that the present method  will also be
applicable when the potential is time-dependent. In that case, we
simply replace all $V({\bf x})$, $V({\bf x}^\prime)$, $V_0 ({\bf
x})$ and $V_0 ({\bf x}^\prime)$ in Theorem 1.1 and 1.3
respectively by $V(t, {\bf x})$, $V(t^\prime, {\bf x}^\prime)$,
$V_0 (t, {\bf x})$, and $V_0 (t^\prime, {\bf x}^\prime)$. Also, we
may generalize the present method by replacing $K(t, {\bf x})$ by
$K \left( t, {\bf x}, \mbox{\boldmath $\partial$}, \partial_t
\right)$ which may depend upon both space and time derivatives,
$\mbox{\boldmath $\partial$}$ and $\partial_t \left( \equiv
{\partial \over
\partial t}\right)$.

\section{Inverse Quadratic Potential}

\setcounter{equation}{0}

In this section, we assume that the potential $V({\bf x})$ is a
homogenous function of ${\bf x}$ of degree $-2$, i.e., it
satisfies a identity

\begin{equation}
V (\lambda {\bf x}) = {1 \over \lambda^2}\ V({\bf x})
\label{eq:twoone}
\end{equation}

\ni for any non-zero real number $\lambda$. For example, we may
assume

$$V({\bf x}) = {\alpha \over x^2_1 + x^2_2 + \cdots + x^2_n} +
\sum^n_{j=1} \ {a_j \over \left( x_j \right)^2} + \sum^n_{j,k=1}\
{b_{jk} \over \left( x_j -x_k \right)^2}$$

\ni etc. for some constants $\alpha$, $a_j$ and $b_{jk}$.

We must choose $f_j (t) =0$ with $x^\prime_j = \xi (t) x_j$ in
this case. Then Eq. (\ref{eq:twoone}) implies the validity of

$$V({\bf x}) = \xi^2 (t) V({\bf x}^\prime ) = \dot \phi (t) V({\bf
x}^\prime )$$

\ni for $\lambda = \xi (t)$ so that Eq. (\ref{eq:oneeighteen})
becomes

$$\left\{ {\partial \over \partial t} - k \Delta \right\} K(t,
{\bf x}) =0$$

\ni which gives

\begin{eqnarray*}
\dot A (t) &=& 2nk C(t) \quad , \\
\noalign{\vskip 4pt}%
\dot C (t) &=& 4k (C(t))^2 \quad .
\end{eqnarray*}

\ni Solving these with Eqs. (1.17)  for $\xi (t)$ and $\phi (t)$,
we find:

\bigskip

\ni \underbar{\bf Proposition 2.1}

\medskip

For any function $\psi (t,{\bf x})$ satisfying

\begin{equation}
{\partial \over \partial t}\  \psi (t, {\bf x}) = k\{ \Delta - V
({\bf x})\} \psi (t, {\bf x}) \label{eq:twotwo}
\end{equation}

\ni with the condition Eq. (\ref{eq:twoone}) for $V({\bf x})$, a
new function given by

\begin{equation}
\psi^\prime (t, {\bf x}) = \left( {1 \over at +b} \right)^{{n
\over 2}} \exp \left\{ -{a \over 4k (at+b)}\ {\bf x}^2 \right\}
\psi (t^\prime , {\bf x}^\prime ) \label{eq:twothree}
\end{equation}

\ni also satisfies  the same equation, i.e.,

\begin{equation}
{\partial \over \partial t} \ \psi^\prime (t, {\bf x}) = k \{
\Delta -V({\bf x})\} \psi^\prime (t , {\bf x}) \quad ,
\label{eq:twofour}
\end{equation}

\ni where $t^\prime$ and ${\bf x}^\prime$ are given by

\begin{equation}
t^\prime = {ct+d \over at+b} \quad , \quad {\bf x}^\prime = {1
\over at+b}\ {\bf x} \quad . \label{eq:twofive}
\end{equation}

\ni Here, $a$, $b$, $c$, and $d$ are arbitrary real constants
satisfying the condition

\begin{equation}
bc - ad = 1 \quad . \quad \vrule height 0.9ex width 0.8ex depth
-0.1ex
\end{equation}

\bigskip

We will show next that the admissible pair $(U,W)$ of section 1
will offer infinite dimensional realizations of the $SL(2,R)$
group. For this purpose, it is more convenient to consider $2
\times 2$ $SL(2,R)$ matrix $M$ by

\begin{equation}
M = \left( \begin{array}{cc} c &d \\
\noalign{\vskip 6pt}%
a &b \end{array}\right) \quad , \quad \det M =1 \quad .
\label{eq:twoseven}
\end{equation}

\ni Moreover, we collectively write the coordinates as

\begin{equation}
Z = \{ t, {\bf x}\} \label{eq:twoeight}
\end{equation}

\ni on which $M$ is assumed to act as

\begin{equation}
MZ = M \{ t, {\bf x}\} = \{ t^\prime , {\bf x}^\prime \} = \left\{
{ct +d \over at+b} \ ,\ {{\bf x} \over at+b} \right\} \quad .
\label{eq:twonine}
\end{equation}

\ni Since $K(t, {\bf x})$ depends upon parameters $a$, $b$, $c$,
and $d$, we will write it as $K (t, {\bf x}|M)$ so that

\begin{equation}
K(t, {\bf x}|M) \equiv K(Z|M) = \left( {1 \over at+b}\right)^{{n
\over 2}} \exp \left\{ -{a \over 4k (at+b)}\ {\bf x}^2 \right\}
\quad . \label{eq:twoten}
\end{equation}

\ni It is then easy to verify.

\bigskip

\ni \underbar{\bf Proposition 2.2}

\medskip

For any two $SL(2,R)$ matrices $M$ and $M^\prime$, we have

\begin{subequations}\label{foo}
\begin{eqnarray}
&{\rm (i) }& \quad M (M^\prime Z ) = (MM^\prime )Z \label{foo-a}
\label{eq:twoelevena}\\
 \noalign{\vskip 8pt}%
&{\rm (ii)}&\quad K (Z|M^\prime) K(M^\prime Z |M) = K(Z|MM^\prime
) \label{foo-b} \label{eq:twoelevenb}
\end{eqnarray}
\end{subequations}

\ni where $MM^\prime$ implies the standard matrix product of $M$
and $M^\prime$. \ $\vrule height 0.9ex width 0.8ex depth -0.1ex$

\bigskip

The linear operators $U$ and $W$ introduced in section 1 also
depend  upon $M$. However, it is more convenient to rewrite them
as $U \left( M^{-1}\right)$ and $W \left( M^{-1}\right)$ instead
of $U(M)$ and $W(M)$ by the reason which will become clear
shortly, so that Eq. (\ref{eq:twothree}) is rewritten as

\begin{equation}
U \left( M^{-1} \right) \psi (Z) = K (Z|M) \psi (MZ) \quad .
\label{eq:twotwelve}
\end{equation}

\ni Note that

$$M^{-1} = \left( \begin{array}{cc} b &-d \\
\noalign{\vskip 6pt}%
-a &c \end{array}\right) \quad {\rm for} \quad M = \left(
\begin{array}{cc} c &d\\
\noalign{\vskip 6pt}%
 a &b \end{array} \right) \quad .$$

\ni Then, Eqs. (2.11) are immediately translated into the
following:

\bigskip

\ni \underbar{\bf Proposition 2.3}

\medskip

Linear operators $U(M)$ and $W(M)$ satisfy

\begin{subequations}\label{foo}
\begin{eqnarray}
U(MM^\prime ) &=& U(M) U(M^\prime) \quad , \label{foo-a}
\label{eq:twothirteena}\\
\noalign{\vskip 6pt}%
W(MM^\prime) &=& W(M) W(M^\prime) \quad . \label{foo-b}
\label{eq:twothirteenb}
\end{eqnarray}
\end{subequations}

\ni In other words, they offer (infinite dimensional) realizations
of the $SL(2,R)$ group. \ $\vrule height 0.9ex width 0.8ex depth
-0.1ex$

\bigskip

In ending this section, it may be of interest to note the
following:

\bigskip

\ni \underbar{\bf Remark 2.4}

\medskip

The transformation, Eqs. (\ref{eq:twofive}) contain both time
translation and dilatation as special cases. If we choose

\[
M = \left( \begin{array}{cc} 1 &\lambda\\
\noalign{\vskip 6pt}%
0&1\end{array}\right) \quad ,
\]

\ni then Eq. (\ref{eq:twofive}) gives the time translation,

$$t \rightarrow t^\prime = t + \lambda \quad , \quad {\bf x}
\rightarrow {\bf x}^\prime =  {\bf x} \quad .$$

\ni On the other side, the choice of

\begin{subequations}\label{foo}
\begin{eqnarray}
& &M = \left( \begin{array}{cc} c &0 \\
\noalign{\vskip 6pt}%
0 &{1 \over c} \end{array}\right) \quad , \quad (c \not= 0)
\label{foo-a}\label{eq:twofourteena}\\
\noalign{\vskip 6pt}%
\noalign{\hbox{leads to the dilatation}}
 \noalign{\vskip 6pt}%
& &t \rightarrow t^\prime = c^2 t \quad , \quad {\bf x}
\rightarrow {\bf x}^\prime = c {\bf x} \quad . \quad \vrule height
0.9ex width 0.8ex depth -0.1ex \label{foo-b}
\label{eq:twofourteenb}
\end{eqnarray}
\end{subequations}

\bigskip

\ni \underbar{\bf Remark 2.5}

\medskip

For some subgroup of $SL(2,R)$, the linear operator $U(M)$ may
possess a non-trivial fixed point in the function space. Consider
one-dimensional case of $n=1$ with $V ({\bf x}) = 0$. The Jacobi's
theta function $\theta_1 (x|t)$ is given$^{6)}$ by

\begin{equation}
\theta_1 (x|t) = i \sum^\infty_{n=-\infty} (-1)^n \exp \left\{ i
\pi \left( n - {1 \over 2} \right)^2 t + i \pi (2n-1) x \right\}
\label{eq:twofifteen}
\end{equation}

\ni which satisfies one-dimensional Schr\"odinger equation

\begin{equation}
4 \pi i \ {\partial \over \partial t}\ \theta_1 (x|t) =
{\partial^2 \over \partial x^2}\ \theta_1 (x|t) \label{twosixteen}
\end{equation}

\ni with $k= -i/4 \pi$. Moreover, if $a$, $b$, $c$, $d$ are all
integers with $bc -ad =1$, we then have the identity$^{6)}$

\begin{equation}
\theta_1 \left( {x \over at+b} \bigg| {ct +d \over at+b} \right) =
\epsilon (at+b)^{{1 \over 2}} \exp \left( {i \pi ax^2 \over at+b}
\right) \theta_1 (x|t) \label{eq:twoseventeen}
\end{equation}

\ni where $\epsilon$ is a constant satisfying $\epsilon^8 =1$,
whose particular value depends upon the specification of branches
of $(at+b)^{{1 \over 2}}$ in the complex $t$-plane. Setting
$\epsilon =1$. Eq. (2.17) is rewritten as

\begin{equation}
U \left( M_0 \right) \theta_1 (x|t) = \theta_1 (x|t)
\label{eq:twoeighteen}
\end{equation}

\ni for any modular matrix $M_0$. Therefore, the $SL(2,R)$ orbit
of $\theta_1 (x|t)$ is the homogeneous space

$$SL(2,R) /SL(2,Z)$$

\ni where $SL(2,Z)$ is the modular subgroup of $SL(2,R)$ in which
all $a$, $b$, $c$, and $d$ are integers.

\bigskip

\ni \underbar{\bf Remark 2.6}

\medskip

For the\, one-dimensional\, case with $V({\bf x}) =0$,\, the
symmetry group\, is actually a

 \vspace{-.41in}

\ni larger one of $SL(2,R) \ \ \ \,
\begin{picture}(40,40)
\put(-1,3){\circle{11}}\hspace{-7pt} s
\end{picture}
 \!\!\!\!\!\!\!\!\!\!\!\!\!\! T_2(R)$
which is a semi-direct product of $SL(2,R)$ with a two-dimensional
translation group $T_2(R)$. This will be shown as a special case
of $\alpha = \beta = 0$ in Eqs. (3.5)-(3.10) in the next section.

\section{Linear Potential}

\setcounter{equation}{0}

In this section, we will consider the case of the linear
potential. For a while, we restrict ourselves to one-dimensional
space and set

\begin{equation}
V(x) = \alpha + \beta x \label{eq:threeone}
\end{equation}

\ni for constants $\alpha$ and $\beta$. Dropping all sub-indices
such as $j$'s in $f_j (t)$, $B_j (t)$ etc. (since $n=1$), Eqs.
 (\ref{eq:oneeighteen}) and (\ref{eq:onesixteenc}) then gives
 (with $\dot A(t) \equiv {d \over dt} \ A(t)$ etc.)

\begin{subequations}\label{foo}
\begin{eqnarray}
&{\rm (i)}&\quad \dot C (t) = 4k C^2 (t) \quad , \label{foo-a}
\label{eq:threetwoa}\\
\noalign{\vskip 8pt}%
&{\rm (ii)}&\quad \dot B (t) = 4k B(t) C(t) + k \beta \left[ \xi^3
(t) -1 \right] \quad , \label{foo-b}\label{eq:threetwob}\\
\noalign{\vskip 8pt}%
&{\rm (iii)}& \quad \dot A (t) = k \left\{ B^2 (t) + 2 C(t)
\right\} + k \alpha \left( \xi^2 (t) - 1 \right) + k \beta \xi^2
(t) f(t) \quad . \label{foo-c} \label{eq:threetwoc}
\end{eqnarray}
\end{subequations}

\ni Together with Eqs. (1.17) which give $2k B (t) = \dot f(t)/\xi
(t)$ and $4k C(t) = \dot \xi (t) /\xi (t)$, we can solve these
equations. In this case, the general solution contains five
independent real parameters. It is convenient for our purpose to
parametrize them as

\begin{subequations}\label{foo}
\begin{eqnarray}
\Lambda &=& \left\{ M, \left( \begin{array}{c} \mu\\
\noalign{\vskip 6pt}%
\nu \end{array}\right) \right\}
\label{foo-a}\label{eq:threethreea}\\
\noalign{\vskip 6pt}%
M &=& \left( \begin{array}{cc} c & d\\
\noalign{\vskip 6pt}%
a &b \end{array}\right) \quad , \quad \det M = 1 \label{foo-b}
\label{eq:threethreeb}
\end{eqnarray}
\end{subequations}

\ni where $M$ is the real $2 \times 2$ unimodular matrix just as
in the previous section, which acts now on 2-dimensional real
vector $\left( \begin{array}{c} \mu\\ \noalign{\vskip 4pt} \nu
\end{array}\right)$ in the parameter space. We write $C(t)$,
$B(t)$ etc. now as $C(t|\Lambda)$, $B(t|\Lambda)$ etc. in order to
indicate their dependence on parameters involved in $\Lambda$. We
also rewrite $K(t,x)$ of section 1 as

\begin{equation}
K(t,x|\Lambda ) = {1 \over \sqrt{at+b}} \exp \left\{ A(t|\Lambda )
+ B(t|\Lambda) x + C (t|\Lambda) x^2 \right\} \quad ,
\label{eq:threefour}
\end{equation}

\ni where we changed however $A(t)$ there into $A(t|\Lambda) - {1
\over 2} \log (at+b)$ for simplicity.  Then, their explicit forms
are found to be

\begin{subequations}\label{foo}
\begin{eqnarray}
&{\rm (i)}& \quad C(t|\Lambda) = -{1 \over 4k}\ {a \over at+b}
\quad , \label{foo-a} \label{threefivea}\\
\noalign{\vskip 8pt}%
&{\rm (ii)}& \quad B(t|\Lambda ) = - {\nu \over 2k}\ {1 \over
at+b} + {k \beta \over 2}\ \left\{ {2 (ct+d) \over (at+b)^2} - t -
{bt \over at +b} \right\} \quad , \label{foo-b}
\label{eq:threefiveb}\\
\noalign{\vskip 8pt}%
&{\rm (iii)}&\quad  A(t|\Lambda ) = -{1 \over 4k}\ \mu \nu +
\alpha k \left( {ct+d \over at+b} - t \right) + {\nu^2 \over 4k}\
{ct+d \over at+b} \nonumber\\
\noalign{\vskip 8pt}%
& & \qquad\quad + k \beta \left\{ \mu \ {ct+d \over at+b} - \nu
\left[ \left( {ct+d \over at+b} \right)^2 - {1 \over 2}\ {t^2
\over at+b} \right]
 \right\} \nonumber\\
\noalign{\vskip 8pt}%
& & \qquad\quad + k^3 \beta^2 \left\{ {2 \over 3}\ \left( {ct+d
\over at+b}\right)^3 + {1 \over 12}\ t^3 + {b \over 4}\ {t^3 \over
at+b} - {t^2 (ct+d) \over (at+b)^2}\right\} \quad . \label{foo-c}
\label{eq:threefivec}
\end{eqnarray}
\end{subequations}

\ni In Eq. (\ref{eq:threefivec}), the constant term $-\mu \nu /
4k$ has been added to simplify the expression of $\omega (\Lambda
, \Lambda^\prime )$ given in Eq. (3.16) shortly. We then have the
following Proposition.

\bigskip

\ni \underbar{\bf Proposition 3.1}

\medskip

For any function $\psi (t,x)$ satisfying

\begin{equation}
{\partial \over \partial t}\ \psi (t,x) = k \left\{ {\partial^2
\over \partial x^2} - \alpha - \beta x \right\} \psi (t,x) \quad ,
\label{eq:threesix}
\end{equation}

\ni the new wave function given by

\begin{equation}
\psi^\prime (t,x) = K (t,x|\Lambda) \psi (t^\prime , x^\prime )
\label{eq:threeseven}
\end{equation}

\ni satisfies the same, i.e.,

\begin{equation}
{\partial \over \partial t} \ \psi^\prime (t,x) = k \left\{
{\partial^2 \over \partial x^2} - \alpha - \beta x \right\}
\psi^\prime (t,x) \quad , \label{eq:threeeight}
\end{equation}

\ni where $t^\prime$ and $x^\prime$ are defined by

\begin{subequations}\label{foo}
\begin{eqnarray}
t^\prime &=& {ct+d \over at+b} \quad , \label{foo-a}
\label{eq:threeninea}\\
\noalign{\vskip 6pt}%
x^\prime &=& {x \over at+b} + \mu - \nu \ {ct+d \over at+b} + k^2
\beta \left\{ \left( {ct+d \over at+b} \right)^2 - {t^2 \over
at+b} \right\} \quad . \quad \vrule height 0.9ex width 0.8ex depth
-0.1ex \label{foo-b} \label{eq:threenineb}
\end{eqnarray}
\end{subequations}

Next we will  show that the underlying symmetry group is now the
semi-direct product

\begin{equation}
G = SL(2,R) \ \ \ \,
\begin{picture}(40,40)
\put(-1,3){\circle{11}}\hspace{-7pt} s
\end{picture}
 \!\!\!\!\!\!\!\!\!\!\!\!\!\!
 T_2 (R) \quad . \label{eq:threeten}
\end{equation}

\ni Let $\Lambda^\prime \ \in\ G$ with

$$\Lambda^\prime = \left\{ M^\prime , \left( \begin{array}{c}
\mu^\prime \\
\noalign{\vskip 6pt}%
\nu^\prime \end{array}\right) \right\} \quad , \quad \det M^\prime
= 1$$

\ni be the second generic element of $G$ in addition to $\Lambda$
given by Eqs. (3.3). We introduce the product $\Lambda \circ
\Lambda^\prime$ by

\begin{equation}
\Lambda \circ \Lambda^\prime = \left\{ M M^\prime , \left(
\begin{array}{c} \mu \\
\noalign{\vskip 6pt}%
\nu \end{array}\right) + M \left( \begin{array}{c} \mu^\prime \\
\noalign{\vskip 6pt}%
\nu^\prime \end{array} \right) \right\} \label{eq:threeeleven}
\end{equation}

\ni which can easily be verified to be associative and defines the
desired group product of the group $G$. Note that the unit element
1 and the inverse $\Lambda^{-1}$ are then given by

\begin{subequations}\label{foo}
\begin{eqnarray}
1 &=& \left\{ \left( \begin{array}{cc} 1 &0\\
\noalign{\vskip 6pt}%
0 &1 \end{array}\right) \ \ ,\ \ \left( \begin{array}{c} 0\\
\noalign{\vskip 6pt}%
0 \end{array}\right) \right\} \quad ,
\label{foo-a}\label{eq:threetwelvea}\\
\noalign{\vskip 6pt}%
\Lambda^{-1} &=& \left\{ M^{-1} , -M^{-1} \left( \begin{array}{c}
\mu \\
\noalign{\vskip 6pt}%
\nu \end{array}\right) \right\} \quad , \label{foo-b}
\label{eq:threetwelveb}
\end{eqnarray}
\end{subequations}

\ni respectively. Again, it is convenient to write

\begin{equation}
Z = \{ t,x\} \label{eq:threethirteen}
\end{equation}

\ni collectively for coordinates and assume the action of $\Lambda
\ \in \ G$ to $Z$ to be given by

\begin{equation}
\Lambda Z = \Lambda \{ t,x\} = \{ t^\prime , x^\prime \} \quad ,
\label{eq:threefourteen}
\end{equation}

\ni in terms of $t^\prime$ and $x^\prime$ given by Eqs. (3.9). We
then find

\bigskip

\ni \underbar{\bf Proposition 3.2}

\medskip

We have

\begin{subequations}\label{foo}
\begin{eqnarray}
&{\rm (i)}&\quad \Lambda \{ \Lambda^\prime Z \} = (\Lambda \circ
\Lambda^\prime )Z \label{foo-a} \label{eq:threefifteena}\\
\noalign{\vskip 6pt}%
&{\rm (ii)}&\quad K (Z |\Lambda^\prime )K (\Lambda^\prime
Z|\Lambda) = \exp \{ \omega (\Lambda , \Lambda^\prime )\}
K(Z|\Lambda \circ \Lambda^\prime ) \quad . \label{foo-b}
\label{eq:threefifteenb}
\end{eqnarray}
\end{subequations}

\ni Here, $\omega (\Lambda , \Lambda^\prime)$ is a constant given
by

\begin{equation}
\omega (\Lambda, \Lambda^\prime ) = {1 \over 4k}\ \{ (\mu a - \nu
c) \mu^\prime + (\mu b - \nu d)\nu^\prime \} \label{threesixteen}
\end{equation}

\ni which satisfies the cycle condition

\begin{subequations}\label{foo}
\begin{eqnarray}
\omega \left( \Lambda_1 , \Lambda_2 \right) + \omega \left(
\Lambda_1 \circ \Lambda_2, \Lambda_3 \right) &=& \omega \left(
\Lambda_2 , \Lambda_3 \right) + \omega \left( \Lambda_1 ,
\Lambda_2 \circ \Lambda_3 \right) \label{foo-a}
\label{eq:threeseventeena}\\
\noalign{\vskip 6pt}%
\noalign{\hbox{as well as}}
\noalign{\vskip 6pt}%
\omega \left( \Lambda^{-1}_2 , \Lambda_1^{-1} \right) &=& - \omega
\left( \Lambda_1 , \Lambda_2 \right) \label{foo-b}
\label{eq:threeseventeenb}
\end{eqnarray}
\end{subequations}

\ni for $\Lambda_j\ \in\ G\ (j=1,2,3)$.\ $\vrule height 0.9ex
width 0.8ex depth -0.1ex$

\bigskip

The proof of this Proposition requires unfortunately  long
computations, although it is straightforward. First, we  set

$$\Lambda^\prime Z = \Lambda^\prime \{ t,x\} \equiv \{ \overline
t, \overline x \}$$

\ni so that

\begin{eqnarray*}
\overline t &=& \phi (t|\Lambda^\prime) = {c^\prime t+d^\prime
\over a^\prime t+b^\prime} \quad ,\\
\noalign{\vskip 4pt}%
\overline x &=& \xi (t|\Lambda^\prime ) x + f (t|\Lambda^\prime )
\end{eqnarray*}

\ni with

\begin{eqnarray*}
\xi (t |\Lambda^\prime ) &=& {1 \over a^\prime t + b^\prime} \quad
, \\
\noalign{\vskip 4pt}%
f (t|\Lambda^\prime ) &=& \mu^\prime - \nu^\prime \ {c^\prime t +
d^\prime \over a^\prime t + b^\prime} + k^2 \beta \left\{ \left(
{c^\prime t+d^\prime \over a^\prime t+b^\prime} \right)^2 - {t^2
\over a^\prime t+ b^\prime} \right\} \quad .
\end{eqnarray*}

\ni Since

$$K(Z|\Lambda^\prime) = {1 \over \sqrt{a^\prime t + b^\prime}}
\exp \{ A(t| \Lambda^\prime) + B (t|\Lambda^\prime)x + C
(t|\Lambda^\prime )x^2 \}$$

\ni and

$$K(\Lambda^\prime Z |\Lambda ) = {1 \over \sqrt{a \overline t + b}}
\exp \{ A( \overline t| \Lambda) + B (\overline t|\Lambda)
\overline x + C (\overline t|\Lambda )\overline x^2 \} \quad ,$$

\ni we must now evaluate the product

$$K(Z|\Lambda^\prime) K(\Lambda^\prime Z | \Lambda ) =
{1 \over \sqrt{a^{\prime \prime} t + b^{\prime \prime}}} \exp \{
A_0(t) + B_0 (t)x + C_0 (t)x^2 \} \quad ,$$

\ni with

\begin{eqnarray*}
A_0 (t) &=& A (t|\Lambda^\prime ) + A(\overline t |\Lambda ) + B(
\overline t |\Lambda ) f(t |\Lambda^\prime ) + C (\overline t
|\Lambda ) [f(t|\Lambda^\prime )]^2 \quad , \\
\noalign{\vskip 6pt}%
B_0 (t) &=& B(t|\Lambda^\prime ) + B (\overline t |\Lambda ) \xi
(t|\Lambda^\prime ) + 2 C(\overline t |\Lambda ) \xi
(t|\Lambda^\prime) f(t|\Lambda^\prime ) \quad ,\\
\noalign{\vskip 6pt}%
C_0 (t) &=& C(t|\Lambda^\prime ) + C(\overline t|\Lambda ) [ \xi
(t|\Lambda^\prime )]^2 \quad .
\end{eqnarray*}

\ni Using expressions given in Eqs. (3.5), we can then verify the
validity of Eq. (\ref{eq:threefifteenb}) after long calculations.
For its computation, the following identities are however quite
useful to simplify the proof. Let us set

\begin{equation}
M^{\prime \prime} = MM^\prime = \left( \begin{array}{cc} c &d\\
\noalign{\vskip 6pt}%
a &b \end{array} \right) \left( \begin{array} {cc} c^\prime
&d^\prime\\
\noalign{\vskip 6pt}%
a^\prime &b^\prime \end{array}\right) = \left(
\begin{array} {cc}
c^{\prime \prime} & d^{\prime \prime} \\
\noalign{\vskip 6pt}%
a^{\prime \prime} &b^{\prime\prime}\end{array}\right)
\label{eq:threeeighteen}
\end{equation}

\ni for $M$, $M^\prime\ \in\ SL(2,R)$. We then find

\begin{subequations}\label{foo}
\begin{eqnarray}
&{\rm (i)}& \quad a \left( {c^\prime t + d^\prime \over a^\prime t
+ b^\prime}\right) + b = {a^{\prime \prime} t + b^{\prime \prime}
\over a^\prime t + b^\prime} \quad , \quad c \left( {c^\prime t +
d^\prime \over a^\prime t + b^\prime}\right) + d = {c^{\prime
\prime} t + d^{\prime \prime} \over a^\prime t + b^\prime} \quad ,
\label{foo-a} \label{threenineteena}\\
\noalign{\vskip 8pt}%
&{\rm (ii)}& \quad {a \over (a^\prime t + b^\prime ) (a^{\prime
\prime} t + b^{\prime \prime})} = {a^{\prime \prime} \over
a^{\prime \prime} t + b^{\prime \prime}} - {a^\prime \over
a^\prime t + b^\prime} \quad ,
\label{foo-b}\label{threenineteenb}\\
\noalign{\vskip 8pt}%
&{\rm (iii)}& \quad a^\prime t + b^\prime = c (a^{\prime \prime} t
+ b^{\prime \prime} ) - a (c^{\prime \prime} t + d^{\prime
\prime})
\quad , \nonumber \\
\noalign{\vskip 8pt}%
& & \quad c^\prime t + d^\prime = - d (a^{\prime \prime} t +
b^{\prime \prime} ) + b (c^{\prime \prime} t + d^{\prime \prime} )
\quad . \label{foo-c} \label{eq:threenineteenc}
\end{eqnarray}
\end{subequations}

\ni Also, for the proof at Eqs. (3.17), it is more convenient to
rewrite Eq. (3.16) in the matrix notation of

\begin{equation}
4 k \omega (\Lambda , \Lambda^\prime ) = \left( \begin{array}{c}
\mu\\
\noalign{\vskip 6pt}%
\nu \end{array}\right)^T J M \left( \begin{array}{c} \mu^\prime\\
\noalign{\vskip 6pt}%
\nu^\prime \end{array}\right) \label{eq:threetwenty}
\end{equation}

\ni where $\left( \begin{array}{c} \mu \\
\noalign{\vskip 4pt} \nu \end{array}\right)^T$ is the transpose
of $\left( \begin{array}{c} \mu\\ \noalign{\vskip 4pt}%
\nu \end{array}\right)$, and $J$ is given by

\begin{subequations}
\begin{eqnarray}
J &=& \left( \begin{array}{cc} 0 &1 \\
\noalign{\vskip 6pt}%
-1 &0 \end{array}\right) \label{foo-a}\label{threetwentyonea}\\
\noalign{\vskip 6pt}%
\noalign{\hbox{which satisfies}}
\noalign{\vskip 6pt}%
M^T J M &=& M J M^T = J \label{threetwentoneb}
\end{eqnarray}
\end{subequations}

\ni for any $M\ \in\ SL(2,R)$ and its transpose matrix $M^T$.

Rewriting $U$ and $W$ of section 1 as $U \left( \Lambda^{-1}
\right)$ and $W \left( \Lambda^{-1} \right)$ so that

\begin{equation}
U \left( \Lambda^{-1} \right) \psi (Z) = \psi^\prime (Z) = K
(Z|\Lambda) \psi (Z^\prime ) \quad , \label{threetwentytwo}
\end{equation}

\ni Proposition 3.2 immediately leads to:

\bigskip

\ni \underbar{\bf Corollary 3.3}

\medskip

We have

\begin{subequations}
\begin{eqnarray}
U ( \Lambda \circ \Lambda^\prime ) &=& \exp \{ - \omega (\Lambda ,
\Lambda^\prime )\} U(\Lambda )U (\Lambda^\prime ) \quad ,
\label{foo-a} \label{eq:threetwentythreea}\\
\noalign{\vskip 4pt}%
W (\Lambda \circ \Lambda^\prime ) &=& \exp \{ - \omega (\Lambda ,
\Lambda^\prime )\} W (\Lambda )W (\Lambda^\prime ) \quad .
\label{foo-b} \label{eq:threetwentythreeb}
\end{eqnarray}
\end{subequations}

\ni In other words, both $U(\Lambda )$ and $W( \Lambda)$ offer
projective representations of $G = SL(2,R)$

\vspace{-.41in}

\ni $ \ \ \,
\begin{picture}(40,40)
\put(-1,3){\circle{11}}\hspace{-7pt} s
\end{picture}
 \!\!\!\!\!\!\!\!\!\!\!\!\!\! T_2 (R)$. We note here that the
 cycle condition Eq. (\ref{eq:threeseventeena}) ensures the
 compatibility of Eqs. (3.23) with the associativity of products
 $U(\Lambda) U(\Lambda^\prime )$ and $W(\Lambda )
 W(\Lambda^\prime)$.$\vrule height 0.9ex width 0.8ex depth
-0.1ex$

Before going into further detail, it may be worthwhile to make the
following remark.

\bigskip

\ni \underbar{\bf Remark 3.4}

\medskip

\vspace{-.41in}

The group $G = SL(2,R) \ \ \ \,
\begin{picture}(40,40)
\put(-1,3){\circle{11}}\hspace{-7pt} s
\end{picture}
 \!\!\!\!\!\!\!\!\!\!\!\!\!\! T_2 (R)$ contains, in some sense,
 the Galilean group. Consider a set of all $\Lambda \ \in\ G$ of
 form

\begin{equation}
\Lambda = \left\{ \left( \begin{array}{cc} 1 &\lambda\\
\noalign{\vskip 6pt}%
0 &1 \end{array}\right) \ ,\ \left( \begin{array}{c} \mu\\
\noalign{\vskip 6pt}%
\nu \end{array}\right) \right\} \label{eq:threetwentyfour}
\end{equation}

\ni which causes the coordinate transformation,

\begin{subequations}\label{foo}
\begin{eqnarray}
t &\rightarrow&  t^\prime = t + \lambda \nonumber \\
\noalign{\vskip 4pt}%
x  &\rightarrow&  x^\prime = x + \sigma + vt \label{foo-a}
\label{eq:threetwentyfivea}\\
\noalign{\vskip 6pt}%
\noalign{\hbox{with}}
\noalign{\vskip 6pt}%
\sigma &=& \mu - \nu \lambda + k^2 \beta \lambda^2 \nonumber \\
\noalign{\vskip 4pt}%
v &=& 2 k^2 \beta \lambda - \nu \label{foo-b}
\label{eq:threetwentyfiveb}
\end{eqnarray}
\end{subequations}

\ni by Eqs. (3.9). We note that the classical Newton's equation $m
\ddot{x} = F$ is invariant under \ the \ Galilean\  transformation
since the \ Newtonian force $F$ is\  constant\  for the

\vspace{-.40in}

\ni linear potential. Therefore, in some sense, the group $G =
SL(2,R) \ \ \ \,
\begin{picture}(40,40)
\put(-1,3){\circle{11}}\hspace{-7pt} s
\end{picture}
 \!\!\!\!\!\!\!\!\!\!\!\!\!\! T_2$ together
with  Eqs. (3.9) may be said to be a quantum-mechanical
generalization of the Galilean transform. \ $\vrule height 0.9ex
width 0.8ex depth -0.1ex$

We will next give examples of Theorem 1.3 for the case of $V_0
({\bf x}) = 0$ and $V( {\bf x} ) = \alpha + \beta x$. For
simplicity, we set

\begin{subequations}\label{foo}
\begin{eqnarray}
{\cal H}_0 &=& \left\{ \psi_0 (t,x) \bigg| \left( {\partial \over
\partial t} - k \Delta \right) \psi_0 (t,x) = 0 \right\}
\label{foo-a} \label{eq:threetwentysixa}\\
\noalign{\vskip 4pt}%
{\cal H} &=& \left\{ \psi (t,x) \bigg| \left( {\partial \over
\partial t} - k \Delta + k \alpha + k \beta x \right) \psi (t,x) =
0 \right\} \quad . \label{foo-b} \label{eq:threetwentysixb}
\end{eqnarray}
\end{subequations}

\ni First, we note that Eq. (\ref{eq:onetwelve}) for $V_0 ({\bf
x}) = 0$ implies $K (t,x)\ \in\ {\cal H}$. Solving conditions
stated in Theorem 1.3, we then have two distinct solutions,
corresponding to $C(x) = 0\ {\rm or}\ \not= 0$. Rewriting these
$K(t,x)$ now as $f_j (t,x)\ (j=1,2)$. We have:

\bigskip

\ni \underbar{\bf Proposition 3.5}

\medskip

Let us set

\begin{subequations} \label{foo}
\begin{eqnarray}
f_1 (t,x) &=& \exp \left\{ -k (\alpha + \beta x) t + {1 \over 3}\
k^3 \beta^2 t^3 \right\} \quad , \label{foo-a}
\label{eq:threetwentysevena}\\
\noalign{\vskip 4pt}%
f_2 (t,x) &=& \sqrt{{1 \over t}}\exp \left\{ -kt \left( \alpha +
{\beta \over 2}\ x \right) + {1 \over 12}\ k^3 \beta^2 t^3 - {x^2
\over 4kt}\right\} \quad , \label{foo-b}
\label{eq:threetwentysevenb}
\end{eqnarray}
\end{subequations}

\ni both of which are elements of ${\cal H}$. Then, for any
$\psi_0 (t,x)\ \in\ {\cal H}_0$, the functions defined by

\begin{equation}
\psi_j (t,x) = f_j (t,x) \psi_0 \left( t^\prime_j , x^\prime_j
\right) \ ,\ (j=1,2) \label{eq:threetwentyeight}
\end{equation}

\ni are elements of ${\cal H}$, where

\begin{subequations} \label{foo}
\begin{eqnarray}
t_1^\prime &=& t \quad , \quad x_1^\prime = x - k^2 \beta t^2
\quad , \label{foo-a} \label{eq:threetwentyninea}\\
\noalign{\vskip 4pt}%
t^\prime_2 &=& - {1 \over t} \quad , \quad x_2^\prime = {x \over
t} - k^2 \beta t \quad . \label{foo-b} \label{eq:threetwentynineb}
\end{eqnarray}
\end{subequations}

\ni Conversely, suppose $\psi (t,x)\ \in\ {\cal H}$. Then, new
functions defined by

\begin{equation}
\psi_0^{(j)} (t,x) = \phi_j (t,x) \psi \left( \tilde t_j , \tilde
x_j \right) \ ,\ (j=1,2) \label{eq:threethirty}
\end{equation}

\ni are elements of ${\cal H}_0$, where we have set

\begin{subequations} \label{foo}
\begin{eqnarray}
\phi_1 (t,x) &=& \exp \left\{ k [ \alpha + \beta x] t + {2 \over
3}\ k^2 \beta^3 t^3 \right\}  \label{foo-a}
\label{eq:threethirtyonea}\\
\noalign{\vskip 4pt}%
\phi_2 (t,x) &=& {1 \over \sqrt{t}} \exp \left\{ - {k \alpha \over
t} - {2 \over 3}\ {k^2 \beta^3 \over t^3} - {k \beta x \over t^2}
- {x^2 \over 4kt} \right\} \quad ,  \label{foo-b}
\label{eq:threethirtyoneb}
\end{eqnarray}
\end{subequations}

\ni with

\begin{subequations} \label{foo}
\begin{eqnarray}
\tilde t_1 &=& t \quad , \quad \tilde x_1 = x + k^2 \beta t^2
\label{foo-a} \label{eq:threethirtytwoa}\\
\noalign{\vskip 4pt}%
\tilde t_2 &=& - {1 \over t} \quad , \quad \tilde x_2 = {x \over
t} + {k^2 \beta \over t^2} \quad . \quad \vrule height 0.9ex width
0.8ex depth -0.1ex \label{foo-b} \label{eq:threetwentysevenb}
\end{eqnarray}
\end{subequations}

So  far,  we  have  considered  only  one-dimensional  problems.
However,  we  can  find some \hfill examples \hfill for \hfill
multi-dimensional \hfill cases \hfill with \hfill the \hfill same
symmetry \hfill group \hfill $G =$

\vspace{-.41in}

\noindent $SL(2,R) \ \ \ \,
\begin{picture}(40,40)
\put(-1,3){\circle{11}}\hspace{-7pt} s
\end{picture}
 \!\!\!\!\!\!\!\!\!\!\!\!\!\! T_2 ( R )$.

\bigskip

\ni \underbar{\bf Example 3.6}

\medskip

Suppose that $\psi (t, {\bf x})$ with ${\bf x} = \left( x_1 , x_2
, \dots , x_n \right)$ satisfy

\begin{equation}
{\partial \over \partial t} \ \psi (t, {\bf x}) = k \left\{ \Delta
- \sum^n_{j=1} \left( \alpha + \beta x_j \right) - \sum^n_{j,k=1}
{a_{jk} \over \left( x_j -x_k \right)^2} \right\} \psi (t,{\bf x})
\quad , \label{eq:threethirtythree}
\end{equation}

\ni where $a_{jk}$ with $a_{jj} =0$ are some constants. We also
set

\begin{equation}
\tilde K (t, {\bf x}|\Lambda) = \prod^n_{j=1} K \left( t, x_j
|\Lambda \right) \label{eq:threethirtyfour}
\end{equation}

\ni where $K(t,x|\Lambda)$ is given by Eqs. (\ref{eq:threefour})
and (3.5). Moreover, we consider the transformation

\begin{equation}
\psi (t, {\bf x}) \rightarrow \psi^\prime (t, {\bf x}) = \tilde K
(t, {\bf x}| \Lambda ) \psi (t^\prime , {\bf x}^\prime )
\label{eq:threethirtyfive}
\end{equation}

\ni where $t^\prime$ and $x_j^\prime$ are given by Eqs. (3.9) by
replacing $x$ there by $x^\prime_j$ for each $j=1,2, \dots, n$.
When we note

$$x^\prime_j - x^\prime_k = {1 \over at+b} \ \left( x_j - x_k
\right) \quad ,$$

\ni for $j,k = 1,2,\dots , n$, we can readily verify the validity
of Eq. (\ref{eq:onenine}) so that $\psi^\prime (t, {\bf x})$ is
another solution of Eq. (\ref{eq:threethirtythree}). \ $\vrule
height 0.9ex width 0.8ex depth -0.1ex$

\bigskip

\vfill\eject

\ni \underbar{\bf Example 3.7}

\medskip

As we will see below, the 2-dimensional nonlinear Schr\"odinger
equation (see references 4, and 7-9 on the subject)

\begin{equation}
{\partial \over \partial t} \ \psi \left( t, x_1 , x_2 \right) = k
\left\{ {\partial^2 \over \partial x^2_1} + {\partial^2 \over
\partial x^2_2} + \lambda |\psi \left( t, x_1 , x_2 \right)|^2
\right\} \psi \left( t, x_1 , x_2 \right) \label{threethirtysix}
\end{equation}

\ni possesses also $G = SL(2,R) \ \ \ \,
\begin{picture}(40,40)
\put(-1,3){\circle{11}}\hspace{-7pt} s
\end{picture}
 \!\!\!\!\!\!\!\!\!\!\!\!\!\! T_2(R)$ symmetry in spite of the
 non-linearity of Eq. (3.36), provided that the
 parameter $k$ is purely imaginary. Let $\tilde K \left( t, x_1 ,
 x_2 \right)$ be given again by Eq. (\ref{eq:threethirtyfour})
 with $\alpha = \beta =0$ for $n=2$ so that

\begin{eqnarray}
 \tilde K \left( t, x_1 , x_2 |\Lambda \right) &=& {1 \over at+b}
\exp \bigg\{ -{1 \over 2k}\ \mu \nu + {\nu^2 \over 2k}\ {ct + d
\over at+b}\nonumber \\
& & -  {\nu \over 2k}\ {1 \over at+b}\ \left( x_1 + x_2 \right) -
{1 \over 4k} \ {a \over at+b}\ \left( x^2_1 + x^2_2 \right)
\bigg\} \quad . \label{eq:threethirtyseven}
\end{eqnarray}

\ni We note then that we have

$$| \tilde K \left( t, x_1 , x_2 | \Lambda \right) |^2 = {1 \over
(at +b)^2}$$

\ni if $k$ is purely imaginary. Then, the new function
$\psi^\prime (t, x_1 , x_2 ) $ given by Eq.
(\ref{eq:threethirtyfive}) for $n=2$ also satisfies

$$\left| \psi^\prime \left( t, x_1 , x_2\right) \right|^2 = {1
\over (at+b)^2} \ \left| \psi \left( t^\prime , x^\prime_1 ,
x^\prime_2\right) \right |^2 \quad . $$

\ni As the consequence, it satisfies the analogue of Eq.
(\ref{eq:onenine}), i.e.,

\begin{eqnarray*}
 \bigg\{ {\partial \over \partial t} & - & k \Delta + k \lambda
|\psi^\prime \left( t, x_1 , x_2 \right)|^2 \bigg\} \psi^\prime
\left( t, x_1 , x_2 \right)\\
\noalign{\vskip 4pt}%
& = & {1 \over (at+b)^2}\ \tilde K \left( t,x_1 ,x_2 \right)
\left\{ {\partial \over \partial t^\prime} - k \Delta^\prime + k
\lambda \left| \psi \left( t^\prime , x^\prime_1 , x^\prime_2
\right) \right|^2 \right\} \psi \left( t^\prime, x^\prime_1 ,
x^\prime_2 \right) \\
\noalign{\vskip 4pt}%
& = &  0 \quad . \quad \vrule height 0.9ex width 0.8ex depth
-0.1ex
\end{eqnarray*}

\bigskip

\ni \underbar{\bf Remark 3.8}

\medskip

Let us return now to the function $f_1 (t,x)$ given by Eq.
(\ref{eq:threetwentysevena}). One thing interesting about this
function is that it is intimately related to a bound state
problem. Consider the eigenvalue problem of

\begin{equation}
\left\{ -{d^2 \over dx^2} + \beta x \right\} u(x) = E u (x)
\label{eq:threethirtyeight}
\end{equation}

\ni with the boundary condition $u(0) =0$ at $x=0$ for some
eigenfunction $u(x)\ \in\ L^2 (0, \infty)$. To see the connection,
we first note

\begin{equation}
{\partial \over \partial t}
 f_1 (t,x) = k \left\{ {\partial^2 \over \partial x^2} - (\alpha + \beta x)
  \right\} f_1 (t,x) \label{eq:threethirtynine}
\end{equation}

\ni since $f_1 (t,x)\ \in\ {\cal H}$. We next choose $k=-i$, and
observe

$$\lim_{t \rightarrow \pm \infty} f_1 (t+i \delta , x) = 0 \quad
,$$

\ni for any $\delta > 0$. Therefore, if we set

\begin{equation}
u(x) = \int^\infty_{-\infty} dt\ f_1 (t+i \delta ,x)
\label{eq:threeforty}
\end{equation}

\ni and integrate Eq. (\ref{eq:threethirtynine}) in $t$, $u(x)$
satisfies Eq. (\ref{eq:threethirtyeight}) with $E=-\alpha$.
Moreover, letting $\delta \rightarrow + 0$, we calculate

\begin{equation}
u(x) = 2 \int^\infty_0 dt\ \cos \left\{ (\alpha + \beta x) t + {1
\over 3} \ \beta^2 t^3 \right\} \label{eq:threefortyone}
\end{equation}

\ni which is the Airy's function$^{10)}$ with $u(x)\ \in\ L^2 (0,
\infty)$ for $\beta > 0$. Therefore, if we set $x=0$ with $\alpha
= -E$, the boundary condition $u(0) =0$ leads to

\begin{equation}
\int^\infty_0 dt \cos \left( {1 \over 3}\ \beta^2 t^3 - E t\right)
= 0 \quad . \label{eq:threefortytwo}
\end{equation}

\ni The relevance of this solution to the quarquonium spectra for
the $S$-wave bound states of the 3-dimensional confining linear
potential can be found in ref. 11.

\bigskip

\ni \underbar{\bf Remark 3.9}

\medskip

Another interesting property of the function $f_1 (t,x)$ is that
it is invariant under the following 2-dimensional Abelian
sub-group $G_0$ of $G$, which consists of all elements of form:

\begin{equation}
\Lambda_0 = \left\{ \left( \begin{array}{cc} 1 &\lambda \\
\noalign{\vskip 6pt}%
0 &1 \end{array}\right)\ \ ,\ \  \left( \begin{array}{c} \mu \\
\noalign{\vskip 6pt}%
0\end{array}\right)\right\} \quad . \label{eq:threefortythree}
\end{equation}

\ni Then, it is straightforward to show the validity of

$$U \left( \Lambda_0 \right) f_1 (t,x) = f_1 (t,x)$$

\ni so that the $G$-orbit of $f_1 (t,x)$ is the symmetric space
$G/G_0$. Note that under $\Lambda_0$, the coordinate transform as
a special Galilean transformation of

\begin{eqnarray*}
t \rightarrow t^\prime &=& t + \lambda \quad ,\\
\noalign{\vskip 4pt}%
x \rightarrow x^\prime &=& x + \left( \mu + k^2 \beta \lambda^2
\right) + 2k^2 \beta \lambda t
\end{eqnarray*}

\ni by Eqs. (3.25).

\bigskip

\ni \underbar{\bf Remark 3.10}

\medskip

If we set $\mu = \nu =0$, then the group $G$ reduces to $SL(2,R)$.
Consider now the following time-dependent potential

\begin{equation}
V (t,x) = \alpha + \beta x + {\lambda \over \left( x -k^2 \beta
t^2 \right)^2} \label{eq:threefortyfour}
\end{equation}

\ni for a constant $\lambda$.  Then, the wave function $\psi
(t,x)$ satisfying

\begin{equation}
{\partial \over \partial t}\ \psi (t,x) = k \left\{ {\partial^2
\over \partial x^2} - V (t,x) \right\} \psi (t,x)
\label{eq:threefortyfive}
\end{equation}

\ni is still invariant under the $SL(2,R)$ symmetry, since we will
have

\begin{equation}
x^\prime - k^2 \beta t^{\prime 2} = {1 \over at+b}\ \left( x - k^2
\beta t^2 \right) \label{eq:threefortysix}
\end{equation}

\ni under Eqs. (3.9) with $\mu = \nu = 0$. In section 5, we will
also show that any function $\psi (t,x)$ satisfying Eq. (3.45) is
an eigenstate of the Casimir invariant $I_2$ of the $s \ell (2)$
Lie algebra.

\section{Quadratic Potential}

\setcounter{equation}{0}

The same method given in the previous sections is also applicable
to the case of the quadratic potential

\begin{equation}
{\partial \over \partial t}\ \psi (t,x) = k \left\{ {\partial^2
\over \partial x^2} - \left( \alpha + \omega^2 x^2 \right)
\right\} \psi (t,x) \quad , \label{eq:fourone}
\end{equation}

\ni for real constants $\alpha$ and $\omega^2$. The value of
$\omega^2$ could also assume a negative value in what follows.
However, in order to avoid the question of the reality constraint
for $t^\prime$ and $x^\prime$, we will temporarily suppose that
variables $t$, $x$, $t^\prime$, and $x^\prime$ as well as other
parameters are allowed to assume complex values. We now perform
the coordinate transformation

\begin{subequations}\label{foo}
\begin{eqnarray}
t  \rightarrow t^\prime &=& \phi (t) \quad , \label{foo-a}
\label{eq:fourtwoa}\\
\noalign{\vskip 4pt}%
x \rightarrow x^\prime &=& \xi (t) x + f (t) \label{foo-b}
\label{eq:fourtwob}
\end{eqnarray}
\end{subequations}

\ni as before with

\begin{subequations}\label{foo}
\begin{eqnarray}
\psi (t,x) &\rightarrow& \psi^\prime (t,x) = K(t,x) \psi (t^\prime
,x^\prime)  \label{foo-a} \label{eq:fourthreea}\\
\noalign{\vskip 4pt}%
K(t,x) &=& \exp \{ A(t) +B(t)x +C(t) x^2\} \label{foo-b}
\label{eq:fourthreeb}
\end{eqnarray}
\end{subequations}

\ni as in section 1. Then Eqs. (1.17) and (\ref{eq:oneeighteen})
for $n=1$ give differential equations

\begin{subequations}\label{foo}
\begin{eqnarray}
&{\rm (i)}&\quad \xi (t) \ddot{\xi} (t) - 2 \dot \xi (t) \dot \xi
(t) = 4k^2 \omega^2 \xi^2 (t) \left\{ \xi^4 (t) -1 \right\}
\label{foo-a} \label{eq:fourfoura}\\
\noalign{\vskip 8pt}%
 &{\rm (ii)}&\quad \xi (t) \ddot{f} (t) - 2 \dot \xi (t) \dot f(t) = 4k^2 \omega^2 \xi^5 (t) f(t) \label{foo-b} \label{eq:fourfourb}\\
\noalign{\vskip 8pt}%
&{\rm (iii)}& \quad \dot \phi (t) = \xi^2 (t) \label{foo-c}
\label{eq:fourfourc}
\end{eqnarray}
\end{subequations}

\ni among many others.

First, the solution of Eq. (\ref{eq:fourfoura}) leads to

\begin{equation}
\xi^2 (t) = \pm {\exp (4k\omega t) \over \{ b+a\,  \exp (4 k\omega
t)\} \{ d+c \, \exp (4k \omega t)\}} \label{eq:fourfive}
\end{equation}

\ni for constants $a$, $b$, $c$, and $d$ satisfying $bc -ad = 1$.
At first glance, this appears rather peculiar, since the 2nd order
differential equation, Eq. (\ref{eq:fourfoura}) admits solutions
containing 3 instead of 2 arbitrary parameters. However, in
writing Eq. (\ref{eq:fourfive}), we took  advantage of the
translation invariance of Eq. (\ref{eq:fourfoura}) under

$$t \rightarrow t^\prime = t + \ {\rm constant} \quad ,$$

\ni which adds one more parameter in theory. Then, the general
solution of Eq. (\ref{eq:fourfourb}) is found to be

\begin{equation}
f(t) = \nu \left[ {d+ c \, \exp (4k \omega t) \over b +a \, \exp
(4k \omega t)} \right]^{{1 \over 2}} - \mu \left[ {b+a \, \exp (4k
\omega t) \over d+c\,  \exp (4k \omega t) } \right]^{{1 \over 2}}
\quad , \label{eq:foursix}
\end{equation}

\ni for additional constant $\mu$ and $\nu$. Therefore, the
solution contains 5 parameters which we specify by

\begin{subequations}\label{foo}
\begin{eqnarray}
\Lambda &=& \left\{ M, \left( \begin{array}{c} \mu \\
\noalign{\vskip 6pt}%
\nu \end{array}\right)\right\}  \label{foo-a}
\label{eq:foursevena}\\
\noalign{\vskip 4pt}%
M &=& \left( \begin{array}{cc} c &d\\
\noalign{\vskip 6pt}%
a &b \end{array} \right) \ ,\ {\rm at}\ M=1 \label{foo-b}
\label{eq:foursevenb}
\end{eqnarray}
\end{subequations}

\ni just as Eqs. (3.3). If  we allow  complex values for all these
parameters, then the pre-

\vspace{-.41in}

\ni sent theory remarkably gives the same symmetry group of
$SL(2,C) \ \ \ \,
\begin{picture}(40,40)
\put(-1,3){\circle{11}}\hspace{-7pt} s
\end{picture}
 \!\!\!\!\!\!\!\!\!\!\!\!\!\! T_2 ( C )$ also as we will see below.
However if we restrict ourselves to real values for $t$, $x$,
$t^\prime$, and $x^\prime$, then we will have the following
complications. Because of the square roots operations for $\xi(t)$
in Eq. (4.5) as well as in $f(t)$ of Eq (\ref{eq:foursix}),
$x^\prime$ given by Eq. (\ref{eq:fourtwob}) will not remain real
for arbitrary real values of $a$, $b$, $c$, $d$, $\mu$ and $\nu$.
We will discuss the problem later.

Since all functions $\xi(t)$, $f(t)$ etc. depend upon the
parameters of $\Lambda$, we rewrite them again as $\xi
(t|\Lambda)$, $f(t|\Lambda)$ etc. However, the formulae become
simpler, if we use the new variable

\begin{subequations}\label{foo}
\begin{eqnarray}
u &=& \exp (4k \omega t)  \label{foo-a} \label{eq:foureighta}\\
\noalign{\vskip 4pt}%
u^\prime &=& \exp (4k \omega t^\prime ) \label{foo-b}
\label{eq:foureightb}
\end{eqnarray}
\end{subequations}

\ni instead of $t$ and $t^\prime$.

We can then rewrite Eq. (4.2) as

\begin{subequations}\label{foo}
\begin{eqnarray}
u^\prime &=& {cu +d \over au +b} \quad , \quad bc - ad = 1
\label{foo-a} \label{eq:fourninea}\\
\noalign{\vskip 4pt}%
x^\prime &=& \xi (t|\Lambda )x + f(t|\Lambda) \label{foo-b}
\label{eq:fournineb}
\end{eqnarray}
\end{subequations}

\ni where

\begin{subequations}\label{foo}
\begin{eqnarray}
\xi (t|\Lambda ) &=& \left[ {u \over (au +b)(cu +d)} \right]^{{1
\over 2}} \quad ,  \label{foo-a} \label{eq:fourtena}\\
\noalign{\vskip 4pt}%
 f(t|\Lambda) &=& \nu \left[ {cu +d \over au+b}\right]^{{1 \over 2}} - \mu \left[
 {au+b \over cu+d}\right]^{{1 \over 2}} \quad . \label{foo-b} \label{eq:fourtenb}
\end{eqnarray}
\end{subequations}

\ni Calculating now explicit forms of $A(t|\Lambda)$, $B
(t|\Lambda)$, and $C(t|\Lambda)$ as in the previous section, we
find:

\bigskip

\ni \underbar{\bf Proposition 4.1}

\medskip

For any $\psi (t,x)$ satisfying Eq. (\ref{eq:fourone}), the new
function given by

\begin{equation}
\psi^\prime (t,x) = K(t, x|\Lambda) \psi (t^\prime , x^\prime )
\left(= U\left( \Lambda^{-1}\right) \psi (t,x) \right)
\label{eq:foureleven}
\end{equation}

\ni is also a solution of the same differential equation, Eq.
(\ref{eq:fourone}). Here, we have set

\begin{subequations}\label{foo}
\begin{eqnarray}
K(t,x|\Lambda) &=& \exp \{ A(t|\Lambda) + B(t|\Lambda) x + C
(t|\Lambda)x^2 \} \quad , \label{foo-a} \label{eq:fourtwelvea}\\
\noalign{\vskip 4pt}%
A (t|\Lambda) &=& {1 \over 4}\log {u \over (au+b)(cu+d)} + {\alpha
\over 4 \omega} \left\{ \log \left( {cu +d \over au+b}\right) -
\log u \right\}\nonumber \\
\noalign{\vskip 4pt}%
 &  & + {\omega \over 2}\ \left\{ \nu^2 \ {cu+d \over au+b} - \mu^2\ {au+b \over cu+d}
  \right\} \quad , \label{foo-b} \label{eq:fourtwelveb}\\
\noalign{\vskip 4pt}%
B(t|\Lambda) &=& \omega \left\{ {\nu \sqrt{u} \over au+b} + {\mu
\sqrt{u} \over cu +d} \right\} \quad , \label{foo-c}
\label{eq:fourtwelvec}\\
\noalign{\vskip 4pt}%
C (t|\Lambda) &=& {\omega \over 2} \left\{ -1 + {b \over au+b} +
{d \over cu +d} \right\} \quad . \quad \vrule height 0.9ex width
0.8ex depth -0.1ex \label{foo-d} \label{eq:fourtwelved}
\end{eqnarray}
\end{subequations}

We introduce the product $\Lambda \circ \Lambda^\prime$ for two
$\Lambda$ and $\Lambda^\prime$ again by Eq.
(\ref{eq:threeeleven}), which de-

\vspace{-.41in}

\ni fines the group $G = SL(2,C) \ \ \ \,
\begin{picture}(40,40)
\put(-1,3){\circle{11}}\hspace{-7pt} s
\end{picture}
 \!\!\!\!\!\!\!\!\!\!\!\!\!\! T_2
( C)$ for complex $\Lambda$ and $\Lambda^\prime$. We also assume
the action of $\Lambda$ to the coordinate $Z = \{ t,x\}$ to be
given by Eqs. (3.14) so that

\begin{equation}
\Lambda Z = \Lambda \{ t,x\} = \{ t^\prime , x^\prime \} \quad .
\label{eq:fourthirteen}
\end{equation}

\ni We then discover after some  calculations that the exact
analogue of Proposition 3.2 also holds  valid for the present case
except for the fact that $\omega ( \Lambda , \Lambda^\prime)$
there is now replaced by

\begin{equation}
\omega ( \Lambda , \Lambda^\prime ) \rightarrow \tilde \omega
(\Lambda , \Lambda^\prime) = \omega \{ ( \mu a - \nu c) \mu^\prime
+ (\mu b - \nu a) \nu^\prime \} \quad . \label{eq:fourfourteen}
\end{equation}

\ni For the proof of these facts, the identities Eqs. (3.19) (with
 $t \rightarrow u$) as well as

\begin{subequations}\label{foo}
\begin{eqnarray}
{b \over a^{\prime \prime} u + b^{\prime \prime}}\ {a^\prime u +
b^\prime \over c^\prime u + d^\prime} &=& {1 \over c^\prime u +
d^\prime} - {a \over a^{\prime \prime} u + b^{\prime \prime}}
\label{foo-a} \label{eq:fourfifteena}\\
\noalign{\vskip 4pt}%
 {d \over c^{\prime \prime} u + d^{\prime \prime}} \ {a^\prime u+b^\prime
  \over c^\prime u+d^\prime} &=& {1 \over c^\prime u+d^\prime} - {c \over c^{\prime \prime}
   u + d^{\prime \prime}}\label{foo-b} \label{eq:fourfifteenb}
\end{eqnarray}
\end{subequations}

\ni are useful, although we will not go into detail. However, we
do not understand the reason why both cases of linear and
quadratic potentials give at least formally the identical final
results.

\bigskip

\ni \underbar{\bf Remark 4.2}

\medskip

In contrast to the case of linear potential, the special
transformation Eq. (\ref{eq:threetwentyfour})

\vspace{-.41in}

\noindent   of $G = SL(2,R) \ \ \ \,
\begin{picture}(40,40)
\put(-1,3){\circle{11}}\hspace{-7pt} s
\end{picture}
 \!\!\!\!\!\!\!\!\!\!\!\!\!\! T_2(R)$ with Eq. (4.9)
 does not give the Galilean formula Eqs. (3.25) for the present
 problem. This is, of course, expected since the classical
 Newton's formula $m \ddot{x} = F = \lambda x$ (sor some constant
 $\lambda$) is no longer invariant under the Galilean
 transformation. \ $\vrule height 0.9ex width 0.8ex depth -0.1ex$

 So far we have ignored the question of the reality for variables
 $t$, $x$, $t^\prime$, and $x^\prime$. Let us discuss the problem
 in some details below. Since both constants $k$ and $\omega$ are
 assumed to be either real or purely imaginary, so will be the
 product $k \omega$. Suppose first that $k \omega$ is real. Then,
 $u= \exp (4k \omega t)$ is real and positive for real $t$. The
 condition that both $x^\prime$ and $t^\prime$ are real requires
 that $au+b$ and $cu+d$ be real with $(au+b)(cu +d) > 0$ for any
 $u > 0$. This can be possible in general only in a neighborhood of the
 unit element $E = \left( \begin{array}{cc} 1 &0\\
 \noalign{\vskip 3pt}%
  0 &1 \end{array} \right)$ of the $SL(2,R)$ matrix $M$.
  Moreover, the allowed values for $M$ depend upon the time $t$.
  In other words, the symmetry group of the theory is not in
  general global $SL(2,R)$ group, but is the so-called local group
  (or group germ). Alternatively we may better consider a sub-set of
  $SL(2,R)$ such that all $a$, $b$, $c$, $d$ are non-negative.
  Then, the reality condition for $t^\prime$ and $u^\prime$ are
  readily maintained. However, the inverse matrix $M^{-1}$ does
  not satisfy the requirement, then. In this case, the symmetry is
  not a group but a global semi-group consisting of all
  non-negative matrices in $SL(2,R)$, when $k \omega$ is real.
  Note that $\mu$ and $\nu$ are chosen to be real in the present case
  in order to make $x^\prime$ to be real in
  Eq. (4.9b).

On the other side, suppose now that $k \omega$ is purely
imaginary. Then, we have $|u| = |u^\prime | = 1$. In that case,
instead of the parametrization Eq. (\ref{eq:foursevenb}) for $M$,
we may use the conformal mapping in the complex $u$-plane by

\begin{equation}
u \rightarrow u^\prime = e^{2i \theta}\ {u - \lambda \over 1-
\lambda^* u} \label{eq:foursixteen}
\end{equation}

\ni for real $\theta$ and any complex $\lambda$ with $|\lambda |
\not= 1$.  The condition $|u^\prime |=1$ whenever we have $|u| =1$
is automatically guaranteed by Eq. (\ref{eq:foursixteen}) for
arbitrary complex number $\lambda$. In terms of $\theta$ and
$\lambda$, we can express $a$, $b$, $c$, $d$ as

$$a = - {\lambda^* \over \sqrt{1-|\lambda |^2}}\ e^{i \theta} \quad ,
\quad b = {1 \over \sqrt{1-| \lambda |^2}} \ e^{-i \theta} \quad
,\quad \ \quad\ \ \,
$$

\begin{equation}
 c = {1 \over \sqrt{1 -|\lambda |^2}}\ e^{i \theta} = b^* \quad ,
 \quad d = - {\lambda \over \sqrt{1 -|\lambda |^2}}\ e^{- i
 \theta} = a^* \label{eq:fourseventeen}
\end{equation}

\ni for $|\lambda |< 1$. This especially gives a identity

\begin{equation}
cu + d = (au +b)^* u \label{eq:foureighteen}
\end{equation}

\ni for $|u| =1$. Then, Eq. (\ref{eq:fourtena}) gives the desired
result of the reality constraint of

$$\xi (t|\Lambda ) = {1 \over |au +b|} > 0 \quad .$$

\ni If we next rewrite Eq. (\ref{eq:fourtenb}) as

\begin{equation}
f (t|\Lambda ) = \nu \left[ {(au+b)^* u \over au+b} \right]^{{1
\over 2}} - \mu \left[ {(au+b)u^* \over (au +b)^*} \right]^{{1
\over 2}} \label{eq:fournineteen}
\end{equation}

\ni for $|u| =1$, then the reality of $f(t|\Lambda )$ can also be
maintained as long as we have

\begin{equation}
\mu^* = - \nu \quad . \label{eq:fourtwenty}
\end{equation}

\ni We can verify that both Eq. (\ref{eq:fourtwenty}) and $M =
\left( \begin{array}{cc} b^*, &a^*\\
\noalign{\vskip 4pt}%
a , &b \end{array}\right)$ given by Eq. (\ref{eq:fourseventeen})
remains invariant under the composition law Eq.
(\ref{eq:threeeleven}). In conclusion, if $k \omega$ is purely
imaginary,  the \ symmetry  group  \ of  the \ problem is  a
particular global \ sub-group of

\vspace{-.41in}

\noindent  $G = SL(2,C) \ \ \ \,
\begin{picture}(40,40)
\put(-1,3){\circle{11}}\hspace{-7pt} s
\end{picture}
 \!\!\!\!\!\!\!\!\!\!\!\!\!\! T_2(C)$, where we use the
 parametrization of $SL(2,C)$ and $T_2 (C)$ as in Eqs.
 (\ref{eq:fourseventeen}) and (\ref{eq:fourtwenty}). Especially,
 it contains a group of conformal mappings of transforming the
 unit circle onto itself in the complex $u$-plane.

 Last, we would like to present another example for Theorem 1.3
 for the present problem with $V_0 ({\bf x} ) = 0$. Let us
 set

\begin{subequations}\label{foo}
\begin{eqnarray}
{\cal H}_0 &=& \left\{ \psi_0 (t,x) \bigg| \left( {\partial \over
\partial t} - k\ {\partial^2 \over \partial x^2} \right) \psi_0
(t,x) = 0 \right\} \label{foo-a} \label{eq:fourtwentyonea}\\
\noalign{\vskip 6pt}%
{\cal H}_1 &=& \left\{ \psi (t,x) \bigg| \left( {\partial \over
\partial t} - k \ {\partial^2 \over \partial x^2} - k \left(
\alpha + \omega^2 x^2 \right) \right) \psi (t,x) = 0 \right\}
\label{foo-b} \label{eq:fourtwentyoneb}
\end{eqnarray}
\end{subequations}

\ni Solving conditions given in Theorem 1.3, we then find:

\bigskip

\ni \underbar{\bf Proposition 4.2}

\medskip

For any $\psi_0 (t,x) \ \in\ {\cal H}_0$, the function given by

\begin{equation}
\psi (t,x) = K_0 (t,x) \psi_0 (t^\prime , x^\prime)
\label{eq:fourtwentytwo}
\end{equation}

\ni is an element of ${\cal H}$, where

\begin{subequations}\label{foo}
\begin{eqnarray}
t^\prime &=& - {\sigma^2 \over 4 k \omega}\ {1 \over u + \lambda}
\label{foo-a} \label{eq:fourtwentythreea}\\
\noalign{\vskip 4pt}%
x^\prime &=& \sigma\ {\sqrt{u} \over u + \lambda} \ x - {\sigma
\tau \over 2 \omega}\ {1 \over u + \lambda} \quad , \label{foo-b}
\label{eq:fourtwentythreeb}
\end{eqnarray}
\end{subequations}

\ni with

\begin{subequations}\label{foo}
\begin{eqnarray}
K_0 (t,x) &=& {u^{{1 \over 4}} \over (\lambda +u)^{{1 \over 2}}}
\exp \left\{ A_0 (t) + B_0 (t) x + C_0 (t) x^2 \right\}\quad ,
\label{foo-a} \label{eq:fourtwentyfoura}\\
\noalign{\vskip 4pt}%
A_0 (t) &=& - {\tau^2 \over 4 \omega}\ {1 \over u + \lambda} - k
\alpha t \quad , \label{foo-b} \label{eq:fourtwentyfourb}\\
\noalign{\vskip 4pt}%
B_0 (t) &=& \tau \ {\sqrt{u} \over u + \lambda} \quad ,
\label{foo-c} \label{eq:fourtwentyfourc}\\
\noalign{\vskip 4pt}%
C_0 (t) &=& {\omega (\lambda -u) \over 2 (u + \lambda )} \quad .
\label{foo-d} \label{eq:fourtwentyfourd}
\end{eqnarray}
\end{subequations}

\ni Here, $\sigma$, $\lambda$, $\tau$ are arbitrary constants and

\begin{equation}
u = \exp \{ 4k \omega t \} \quad . \quad \vrule height 0.9ex width
0.8ex depth -0.1ex
\end{equation}

\bigskip

\ni \underbar{\bf Remark 4.3}

\medskip

Unfortunately, the variables $t^\prime$ and $x^\prime$ given by
Eqs. (4.23) can be real only for the case of $k \omega$ being
real. For that case, we can construct solutions of Eq.
(\ref{eq:fourtwentyoneb}) from that of $\psi (t,x)$ satisfying Eq.
(\ref{eq:threetwentysixb}) by combining Eqs. (3.30) and (4.22). \
 $\vrule height 0.9ex width 0.8ex depth -0.1ex$

\bigskip

\section{Lie Algebras and Local Symmetry}

\setcounter{equation}{0}

In the preceding sections, we found that the time-dependent
Schr\"odinger equation for some potentials has global groups as
symmetry  of the theory. However, much larger local symmetry could
emerge, if we consider its Lie algebraic structure as follows.

Let us first set for simplicity

\begin{eqnarray}
K &\equiv& {\partial \over \partial t} - k \left\{ {\partial^2
\over \partial x^2} - V (x) \right\} \label{eq:fiveone}\\
\noalign{\vskip 6pt}%
\noalign{\hbox{and}}
\noalign{\vskip 6pt}%
{\cal H} &=& \left\{ \psi (t,x)|K\psi(t,x) = 0 \right\} \quad .
\label{eq:fivetwo}
\end{eqnarray}

\ni Consider now, as an example, the symmetry group $G = SL(2,R) \
\ \ \,
\begin{picture}(40,40)
\put(-1,3){\circle{11}}\hspace{-7pt} s
\end{picture}
 \!\!\!\!\!\!\!\!\!\!\!\!\!\! T_2(R)$ of section 3 for the linear
 potential. We know that for any $\psi (t,x)\ \in\ {\cal H}$, we
 have $U (\Lambda) \psi (t,x)\ \in \ {\cal H}$ for any $\Lambda \
 \in\ G$. Since $G$ is a Lie group, we can associate a Lie algebra
 $L$ by considering infinitesimal $\Lambda$'s. It is then evident
 that we have

 \begin{equation}
g \psi (t,x)\ \in\ {\cal H} \quad , \quad g\ \in\ L \quad .
\label{eq:fivethree}
\end{equation}

\ni Let ${\tilde U}(L)$ be the universal enveloping algebra of
$L$. Also, we will then have

\begin{equation}
\tilde U(L) \psi (t,x)\ \in\ {\cal H} \label{eq:fivefour}
\end{equation}

\ni whenever $\psi (t,x)$ is a sufficiently smooth function of $t$
and $x$. Since $K$ is invariant under the time translation $t
\rightarrow t^\prime = t + \lambda$ for any constant $\lambda$,
the Lie algebra $L$ always contain a special element

\begin{equation}
D\equiv {\partial \over \partial t}\ \in\ L \label{eq:fivefive}
\end{equation}

\ni so that this implies the validity of

\begin{equation}
D^n \psi (t,x)\ \in\ {\cal H} \label{eq:fivesix}
\end{equation}

\ni for any positive integer $n$. This can be, of course, more
directly verified from $\left[ D^n , K \right] = 0$. For the case
of the linear potential of section 3, the Lie algebra $L$ consists
now of 6 elements (instead of 5 with the additional unit element
1);

\begin{equation}
L = \left\{ L_\pm , L_3 , T_1 , T_2 , 1 \right\}
\label{eq:fiveseven}
\end{equation}

\ni which forms the Abelian-extended Lie algebra of

\begin{equation}
L = s\ell (2) \oplus t (2)  \oplus u(1) \label{eq:fiveeight}
\end{equation}

\ni where $u(1)$ is the extra one-dimensional Abelian algebra in
conformity with the projective representation nature of
$U(\Lambda)$ in Eqs. (3.23). Their explicit forms are easily
calculated from Eqs. (\ref{eq:threefour}), (3.5),
(\ref{eq:threeseven}) and (3.9) to be given by

\begin{subequations}\label{foo}
\begin{eqnarray}
-L_3 &=& t \ {\partial \over \partial t} + {1 \over 2}\ \left( x +
3k^2 \beta t^2 \right)\ {\partial \over \partial x} + k \left(
\alpha + {3 \over 2}\ \beta x \right) t   + {1 \over 2}\ k^3
\beta^2 t^3 + {1 \over 4} \ \  ,
\label{foo-a} \label{eq:fiveninea}\\
\noalign{\vskip 4pt}%
L_+ &=& {\partial \over \partial t} + 2k^2 \beta t \ {\partial
\over \partial x} + k \left( \alpha + \beta x \right) + k^3
\beta^2 t^2 \quad , \label{foo-b} \label{fivenineb}\\
\noalign{\vskip 4pt}%
L_- &=& t^2 \ {\partial \over \partial t} + \left( tx + k^2 \beta
t^3 \right) \ {\partial \over \partial x} + {1 \over 2}\ t +
\alpha k t^2 \nonumber\\ & &  + {1 \over 4}\ k^3 \beta^2 t^4 + {3
\over 2}\ k \beta t^2 x + {1 \over 4k}\ x^2 \quad , \label{foo-c}
\label{eq:fiveninec}
\end{eqnarray}
\end{subequations}

\ni and

\begin{subequations}\label{foo}
\begin{eqnarray}
T_1 &=& {\partial \over \partial x} + k \beta t \quad ,
\label{foo-a} \label{eq:fivetena}\\
\noalign{\vskip 4pt}%
T_2 &=& t\ {\partial \over \partial x} + {1 \over 2k}\ x + {k
\beta \over 2}\ t^2 \quad . \label{foo-b} \label{eq:fivetenb}
\end{eqnarray}
\end{subequations}

\ni They satisfy commutation relations:

\begin{subequations}\label{foo}
\begin{eqnarray}
\left[ L_3 , L_\pm \right] &=& \pm L_\pm \quad , \quad \left[ L_+
, L_- \right] = - 2L_3 \quad , \label{foo-a}
\label{eq:fiveelevena}\\
\noalign{\vskip 4pt}%
\left[ L_3 , T_1 \right] &=& {1 \over 2}\ T_1 \quad , \quad \left[
L_3 , T_2 \right] = - {1 \over 2}\ T_2 \quad ,
\label{foo-b}\label{eq:fiveelevenb}\\
\noalign{\vskip 4pt}%
\left[ L_+ , T_1 \right] &=& \left[ L_- , T_2 \right] = 0 \quad ,
\label{foo-c} \label{eq:fiveelevenc}\\
\noalign{\vskip 4pt}%
\left[ L_+ , T_2 \right] &=& T_1 \quad , \quad \left[ L_-, T_1
\right] = - T_2 \quad , \label{foo-d} \label{eq:fiveelevend}\\
\noalign{\vskip 4pt}%
\left[ T_1 , T_2 \right] &=& {1 \over 2k} \quad .
\label{foo-e}\label{eq:fiveelevene}
\end{eqnarray}
\end{subequations}

\ni Note that $\left[ T_1 , T_2 \right] = {1 \over 2k} \not= 0$,
reflecting the projective representation of Eq. (3.23). We note
that $T_1$ and $T_2$ play the role of creation and annihilation
operators.

Since $W(\Lambda)$ is also a representation of $G$, we can perform
the same analysis to find that the corresponding Lie algebra
$\tilde L$ consisting of

\begin{equation}
\tilde L = \left\{ \tilde L_\pm , \tilde L_3 , \tilde T_1 , \tilde
T_2 , 1 \right\} \label{eq:fivetwelve}
\end{equation}

\ni has the form

\begin{eqnarray}
\tilde L_3 &=& L_3 - 1 \quad , \quad \tilde L_+ = L_+ \quad ,
\quad \tilde L_- = L_- + 2 t \quad , \nonumber\\
\noalign{\vskip 4pt}%
\tilde T_1 &=& T_1 \quad , \quad \tilde T_2 = T_2
\label{eq:fivethirteen}
\end{eqnarray}

\ni with the same commutation relation, Eqs. (5.11). When we write

\begin{equation}
K_1 = {\partial \over \partial t} - k \left\{ {\partial^2 \over
\partial x^2} - \alpha - \beta x \right\} \quad ,
\label{eq:fivefourteen}
\end{equation}

\ni then Eq. (\ref{eq:oneeleven})  becomes

\begin{equation}
\tilde L K_1 = K_1 L \quad . \label{fivefifteen}
\end{equation}

\ni This especially  implies $\left[ L_+ , K_1 \right] = \left[
T_1 , K_1 \right] = \left[ T_2 , K_1 \right] = 0$ and $\left[ L_3
, K_1 \right] = K_1$. Moreover, we can easily find that $K_1$ is
rewritten as a element of $\tilde U(L)$ as

\begin{equation}
K_1 = L_+ - k T^2_1 \label{eq:fivesixteen}
\end{equation}

\ni while the time derivative $D = {\partial \over \partial t}$ is
expressed as

$$D = L_+ - 2 k^2 \beta T_2 - k \alpha \quad .$$

Before going into further details, we note first that the second
order Casimir invariant of the $s\ell (2)$ sub-Lie algebra of $L$
is given by

\begin{equation}
I_2 = L_+ L_- - L^2_3 + L_3 \quad . \label{eq:fiveseventeen}
\end{equation}

\ni In contrast, the larger Lie algebra $L$ possesses not the
second but third order Casimir invariant

\begin{equation}
I_3 = -I_2 + k \left\{ L_3 \left( T_1 T_2 + T_2 T_1 \right) + L_+
T_2 T_2 + L_- T_1 T_1 \right\} \quad . \label{eq:fiveeighteen}
\end{equation}

\ni For our particular form of  generators given by Eqs. (5.9) and
Eq. (5.10), we find that $I_3$ is purely a constant

\begin{equation}
I_3 = {3 \over 16} \label{eq:fivenineteen}
\end{equation}

\ni while $I_2$ is rewritten as

\begin{equation}
I_2 = {3 \over 16} + {1 \over 4k}\ \left( x -k^2 \beta t^2
\right)^2 K_1 \quad . \label{eq:fivetwenty}
\end{equation}

\ni Especially for any function $\psi = \psi (t,x)$ satisfying
$K_1 \psi =0$, we have $I_2 \psi = {3 \over 16}\ \psi$. In this
connection, two special functions $f_1 (t,x)$ and $f_2 (t,x)$
given in Eqs. (3.27) have the following interesting property. They
satisfy

\begin{subequations}\label{foo}
\begin{eqnarray}
L_+ f_1 (t,x) &=& T_1 f_1 (t,x) = 0 \quad , \quad L_3  f_1 (t,x) =
- {1 \over 4}\ f_1 (t,x) \quad , \label{foo-a}
\label{eq:fivetwentyonea}\\
\noalign{\vskip 4pt}%
L_- f_2 (t,x) &=& T_2 f_2 (t,x) = 0 \quad , \quad L_3 f_2 (t,x) =
{1 \over 4}\ f_2 (t,x) \quad , \label{foo-b}
\label{eq:fivetwentyoneb}
\end{eqnarray}
\end{subequations}

\ni Therefore, the function $f_1 (t,x)$ corresponds to the highest
weight state of simultaneous representations of both $s\ell (2)$
and $L$, while $f_2 (t,x)$ plays the role of the lowest weight
state of another representation. They are infinite dimensional and
irreducible but \underbar{not} unitary. Moreover, they satisfy
$K_1 f_j (t,x) =0$ for $j=1,2$ so that we have

\begin{equation}
I_2 f_j (t,x) = I_3 f_j (t,x) = {3 \over 16}\ f_j (t,x) \quad ,
\quad (j=1,2) \quad . \label{eq:fivetwentytwo}
\end{equation}

\ni Also, in view of Eq. (5.20), the wave function $\psi (t,x)$
satisfying Eqs. (3.44) and (3.45) is the eigenstate of $I_2$ with
$I_2 \psi (t,x) = \left\{ {3 \over 16} - {\lambda \over 4}\right\}
\psi (t,x)$.

The same analysis is readily applicable for the quadratic
potential of section 4. In this case, the Lie algebras $L$ and
$\tilde L$ are specified by

\begin{subequations}
\begin{eqnarray}
-L_3 &=& u \ {\partial \over \partial u} + {\alpha \over 4 \omega}
\quad , \label{foo-a} \label{eq:fivetwentythreea}\\
\noalign{\vskip 4pt}%
L_+ &=& {\partial \over \partial u} - {1 \over 2u}\ x\ {\partial
\over \partial x} + \left( {\alpha \over 4 \omega} - {1 \over
4}\right)\ {1 \over u} + {\omega \over 2}\ {x^2 \over u} \quad ,
\label{foo-b} \label{eq:fivetwentythreeb}\\
\noalign{\vskip 4pt}%
L_- &=& u^2\ {\partial \over \partial u} + {1 \over 2}\ ux\
{\partial \over \partial x} + \left( {\alpha \over 4 \omega} + {1
\over 4} \right) \ u + {\omega \over 2}\ u x^2 \quad ,
\label{foo-c} \label{eq:fivetwentythreec}\\
\noalign{\vskip 4pt}%
T_1 &=& {1 \over \sqrt{u}}\ {\partial \over \partial x} - {\omega
\over \sqrt{u}}\ x \quad , \label{foo-d}
\label{eq:fivetwentythreed}\\
\noalign{\vskip 4pt}%
T_2 &=& \sqrt{u}\ {\partial \over \partial x} + \omega \sqrt{u}\
x \quad , \label{foo-e} \label{eq:fivetwentythreee}
\end{eqnarray}
\end{subequations}

\ni with $u = \exp (4k \omega t)$, and

\begin{subequations}\label{foo}
\begin{eqnarray}
\tilde L_3 &=& L_3 \quad , \quad \tilde L_+ = L_+ - {1 \over u}
\quad , \quad \tilde L_- = L_- + u \quad , \label{foo-a}
\label{eq:fivetwentyfoura}\\
\noalign{\vskip 4pt}%
\tilde T_1 &=& T_1 \quad , \quad \tilde T_2 = T_2 \quad .
\label{foo-b} \label{eq:fivetwentyfourb}
\end{eqnarray}
\end{subequations}

\ni They satisfy the same commutation relations. Eqs.
(\ref{eq:fiveelevena}--{\ref{eq:fiveelevend}) while Eqs.
(\ref{eq:fiveelevene}) is now replaced by

\begin{equation}
\left[ T_1 , T_2 \right] = 2 \omega \quad .
\label{eq:fivetwentyfive}
\end{equation}

\ni Writing

\begin{equation}
K_2 = {\partial \over \partial t} - k \left\{ {\partial^2 \over
\partial x^2} - \alpha - \omega^2 x^2 \right\} \quad ,
\label{eq:fivetwentysix}
\end{equation}

\ni the analogues of Eqs. (5.15), and (\ref{eq:fivesixteen}), are
now given by

\begin{subequations}\label{foo}
\begin{eqnarray}
\tilde L K_2 &=& K_2 L \label{foo-a} \label{eq:fivetwentysevena}\\
\noalign{\vskip 4pt}%
K_2 &=& -4k \omega L_3 - {k \over 2} \ \left( T_1 T_2 + T_2 T_1
\right) \label{foo-b} \label{eq:fivetwentysevenb}\\
\noalign{\vskip 4pt}%
D &=& {\partial \over \partial t} = - 4 k \omega L_3 - k \alpha
\quad . \label{foo-c} \label{eq:fivetwentysevenc}
\end{eqnarray}
\end{subequations}

\ni Especially, Eqs. (5.24) and (\ref{eq:fivetwentysevena}) lead
to

$$\left[ L_3 , K_2 \right] = \left[ T_1 , K_2 \right] = \left[
T_2 , K_2 \right] = 0 \quad .$$

\ni For this case, the 2nd order Casimir invariant $I_2$ is still
given by Eq. (\ref{eq:fiveseventeen}), while Eq.
(\ref{eq:fiveeighteen}) for $I_3$ must now be replaced by

\begin{equation}
I_3 = -I_2 + {1 \over 4 \omega}\ \left\{ L_3 \left( T_1 T_2 + T_2
T_1 \right) + L_- T_1 T_1 + L_+ T_2 T_2 \right\} \quad .
\label{eq:fivetwentyeight}
\end{equation}

\ni We  still have the validity of $I_3 = {3 \over 16}$ but Eq.
(\ref{eq:fivetwenty}) is now replaced by

\begin{equation}
I_2 = {3 \over 16} + {1 \over 4k}\ x^2 K_2 \quad .
\label{eq:fivetwentynine}
\end{equation}

\ni Especially, if $\psi (t,x)$ now satisfies

$$\left\{ {\partial \over \partial t} - k \left( {\partial^2 \over
\partial x^2} - \alpha - \omega^2 x^2  - {\lambda \over x^2} \right)
\right\} \psi (t,x) = 0 $$

\ni for some constant $\lambda$, then Eq.
(\ref{eq:fivetwentynine}) now implies

$$I_2 \psi (t,x) = \left( {3 \over 16} - {\lambda \over 4} \right)
\psi (t,x) \quad .$$

There exist relations analogous to Eqs. (5.21). Setting

\begin{subequations}\label{foo}
\begin{eqnarray}
g_1 (t,x) &=& \exp \left\{ k \left( \omega - \alpha \right) t +
{\omega \over 2}\ x^2 \right\} \quad ,
\label{foo-a}\label{fivethirtya}\\
\noalign{\vskip 4pt}%
g_2 (t,x) &=& \exp \left\{ -k \left( \omega + \alpha \right) t -
{\omega \over 2}\ x^2 \right\} \quad , \label{foo-b}
\label{eq:fivethirtyb}
\end{eqnarray}
\end{subequations}

\ni it is easy to verify

\begin{subequations}\label{foo}
\begin{eqnarray}
K_2 g_1 (t,x) &=& L_+ g_1 (t,x) = T_1 g_1 (t,x) = 0 \ \  , \ \ L_3
g_1 (t,x) = - {1 \over 4}\ g_1 (t,x) \ \  , \label{foo-a}
\label{eq:fivethirtyonea}\\
\noalign{\vskip 4pt}%
K_2 g_2 (t,x) &=& L_- g_2 (t,x) = T_2 g_2 (t,x) = 0 \ \  , \ \
L_3 g_2 (t,x) = {1 \over 4}\ g_2 (t,x) \quad . \label{foo-b}
\label{eq:fivethirtyoneb}
\end{eqnarray}
\end{subequations}

\ni Note that $g_2 (t,x)$ given by Eq. (5.30b) corresponds to the
ground state wave function of the familiar harmonic potential. It
is again the lowest weight state of representations of both $s
\ell (2)$ and $L$. Another interesting function is obtained by
setting $\tau = \lambda = 0$, and $\sigma =1$ with appropriate
choice for $\psi_0 (t,x)$ in Proposition 4.3. In this way, the
function

\begin{equation}
g_3 (t,x) = \exp \left\{ -k (\alpha + \omega )t - {\omega \over
2}\ x^2 - \gamma \ {x \over \sqrt{u}} - {\gamma^2 \over 4 \omega}\
{1 \over u} \right\}
\end{equation}

\ni for an arbitrary constant $\gamma$ turns out to be a
simultaneous eigenfunction of $K_2$, $T_2$ and $L_-$ as in

\begin{subequations}\label{foo}
\begin{eqnarray}
&{\rm (i)}& \quad K_2 g_3 (t,x) = 0 \quad ,
\label{foo-a}\label{eq:fivethirtythreea}\\
\noalign{\vskip 4pt}%
&{\rm (ii)}& \quad T_2 g_3 (t,x) = \gamma g_3 (t,x) \quad ,
\label{foo-b} \label{eq:fivethirtythreeb}\\
\noalign{\vskip 4pt}%
&{\rm (iii)}& \quad L_- g_3 (t,x) = {\gamma^2 \over 4 \omega}\ g_3
(t,x) \quad . \label{foo-c}\label{eq:fivethirtythreec}
\end{eqnarray}
\end{subequations}

\ni Especially, Eq. (\ref{eq:fivethirtythreeb}) implies that $g_3
(t,x)$ plays a role of coherent state as in the quantum
optics,$^{12)}$ since $T_2$ may be regarded as the analogue of the
annihilation operator.

\bigskip

\ni \underbar{\bf Acknowledgement}

\medskip

This paper is supported in part by U.S. Department of Energy
Contract no.

\ni DE-FG02-91ER40685.

\bigskip
\bigskip

\vfill\eject

\ni \underbar{\bf References}

\medskip

\begin{enumerate}
\item J. Wu and Y. Alhassid, The potential group approach and
hypergeometric differential equations, J. Math. Phys. {\bf 31},
557-562 (1990).

\item M. Sezgin, A.Y. Verdiyev, and Y.A. Verdiyev, Generalized
P\"oschl-Teller, Toda, Morse potential and $SL(2,R)$ group, J.
Math. Phys. {\bf 39}, 1910-1918 (1998).

\item N.M. Nieto and D.R. Truax, Time-dependent Schr\"odinger
equations having isomorphic symmetry algebras, II. Symmetry
algebra, coherent and squeezed state, J. Math. Phys. {\bf 41},
2753-2767 (2000).

\item H.D. Doebner and G. Goldin, Some remarks on non-linear
quantum mechanics, a talk presented at the 24th International
Colloquium on Group Theoretical Methods in Physics, held at Paris,
July 15-20 (2002).

\item M. Eastwood, Symmetry and differential invariants, a talk
presented at the 6th International Conference on Clifford Algebras
and Their Applications held at Tennessee Technological University,
Cookeville, TN, May 18-25 (2002).

\item A. Erdelyi, W. Magnus, F. Oberhettinger, and F.G. Tricomi,
\underbar{Higher Transcen-} \underbar{dental Functions, Bateman
Manuscript Project Vol. 3}, McGraw Hill (NY) 1955.

\item D. Faddev and L.A. Takhtajan, \underbar{Hamiltonian methods
in the theory of solitions}, (Springer, Berlin 1987).

\item A. Das, \underbar{Integrable Models}, (World Scientific,
Singapore 1989).

\item M.A. Ablowitz and P.A. Clarkson, \underbar{Solitons,
non-linear evolution equations and} \underbar{inverse scattering},
(Cambridge, New York, 1991).

\item M. Abramowitz and I.A. Stegun, \underbar{Handbook of
Mathematical Functions}, Dober Pub., NY (1970), see pp.446-447.

\item A. Das and A.C. Melissinos, \underbar{Quantum Mechanics, A
Modern Introduction}, Gordon and Breach, NY (1986), pp.339-352.

\item L. Mandel and E. Wolf, \underbar{Optical Coherence and
Quantum Optics}, Cambridge Univ. Press, NY (1995), pp.538-540.

\end{enumerate}

\end{document}